 \documentclass{article}
\usepackage{graphicx}
\usepackage{epsfig}
\usepackage{amsfonts}

\usepackage{amsmath}
\usepackage{amssymb}
\usepackage{mathrsfs}
\usepackage{graphicx}
\usepackage{float}
\usepackage{caption}
\usepackage{cite}

 \usepackage{amssymb}

\date{\today}

\newcommand{\insertplot}[5]{\begin{figure}
 \hfill\hbox to 0.05in{\vbox to #5in{\vfill
 \inputplot{#1}{#4}{#5}}\hfill}
 \hfill\vspace{-.1in}
 \caption{#2}\label{#3}
 \end{figure}}
 \newcommand{\inputplot}[3]{
 \special{ps: plotfile #1}
}
\newcounter{fig}   \newcommand{\lbfig}[1]{\refstepcounter{fig}
\label{#1} }

\renewcommand{\t}{\theta}

\newcommand{\f}{\phi}

\newcommand{\ee}{\end{equation}}
\newcommand{\eea}{\end{eqnarray}}
\newcommand{\be}{\begin{equation}}
\newcommand{\bea}{\begin{eqnarray}}

\newcommand{\re}[1]{(\ref{#1})}

\usepackage{xcolor}
\definecolor{mygreen}{HTML}{006E28}

\begin{document}

\title{
\bf
Chains of Boson Stars}

\author{
{\large C. A. R. Herdeiro}$^{\dagger}$,
{\large J.~Kunz}$^{\ddagger}$,
{\large }
  {\large I.~Perapechka}$^{\star}$,
  {\large 	E. Radu}$^{\dagger}$  and
	  {\large 	Y. Shnir}$^{\diamond}$
\\
\\
$^{\dagger}${\small Department of Mathematics, University of Aveiro and CIDMA,
Campus de Santiago, 3810-183 Aveiro, Portugal}
\\
$^{\ddagger}$ {\small  Institute of Physics, University of Oldenburg, Germany
Oldenburg D-26111, Germany}
\\
$^{\star}$ {\small  Department of Theoretical Physics and Astrophysics,
Belarusian State University, Minsk 220004, Belarus}
\\
$^{\diamond}$ {\small  BLTP, JINR, Dubna 141980, Moscow Region, Russia}
}

\date{January 2021}

\maketitle

\begin{abstract}
We study axially symmetric multi-soliton solutions of a complex scalar field theory with a sextic potential,
 minimally
coupled to Einstein's gravity.
These solutions carry no angular momentum and can be classified by the
number of nodes of the scalar field, $k_z$, along the symmetry axis; they are interpreted as  \textit{chains} with  $k_z+1$ boson stars, bound by gravity, but kept apart by repulsive scalar interactions.
 Chains with an \textit{odd} number of constituents
show a spiraling behavior for their ADM mass (and Noether charge) in terms of their angular frequency, similarly to a single fundamental boson star,
as long as the gravitational coupling is small; for larger coupling, however, the inner part of the spiral is replaced by a merging with the fundamental branch of radially excited spherical boson stars.
 Chains with an \textit{even} number of constituents  exhibit a truncated spiral pattern,  with only two  or three branches, ending at a limiting solution
with finite values of ADM mass and Noether charge.
\end{abstract}


\section{Introduction}

Many non-linear physical systems support non-topological solitons, which
represent spatially localized field configurations.
One of the simplest examples in flat space
is given by $Q$-balls, which are particle-like configurations in a model with a complex scalar field possessing
 a harmonic time dependence
and a suitable
self-interaction potential \cite{Rosen,Friedberg:1976me,Coleman:1985ki}.
When $Q$-balls are coupled to gravity, the
so-called \textit{Boson Stars} (BSs) emerge,
which represent solitonic solutions with a topologically trivial and
  globally regular geometry.
The simplest such configurations are static and spherically
symmetric, the scalar field possessing a mass term only, without self-interaction
 \cite{Kaup:1968zz,Ruffini:1969qy}.
These  solutions are usually dubbed \textit{mini}-Boson Stars (mBS),
being regarded as {\it macroscopic quantum states}, which are prevented from gravitationally
collapsing by Heisenberg's uncertainty principle; also, they do not  have a regular flat spacetime limit.

Both $Q$-balls and BSs carry a Noether charge associated with an unbroken continuous global $U(1)$ symmetry.
The charge $Q$ is proportional to the angular frequency of the complex boson field and
represents the boson particle number of the configurations
\cite{Friedberg:1976me,Coleman:1985ki}.

In flat spacetime $Q$-balls exist only within a certain frequency range for the scalar field:
between a maximal value  $\omega_{\rm max}$, which corresponds to the mass of the scalar excitations,
and some lower non-zero critical value  $\omega_{\rm min}$,  that depends on the form of the potential.
On the one hand,  as the frequency $\omega$ is approaching its extremal values,
 both the mass $M$ and the  Noether charge $Q$ of the
configurations diverge. On the other hand, $M$ and  $Q$ attain a minimum at some critical value $\omega_{\rm cr}\in [\omega_{\rm min}, \omega_{\rm max}]$.
Away from  $\omega_{\rm cr}$
the mass and the charge of the configurations increase towards the divergent value at the boundary of the domain. Within $[\omega_{\rm min}, \omega_{\rm cr}]$ the configurations become
unstable against decay. 

The situation is different for BSs: gravity modifies the critical
behavior pattern of the configurations.
The fundamental branch of the BS solutions starts off from the perturbative excitations at $\omega \sim\omega_{\rm max}$,  at which $M,Q$ trivialise (rather than diverge).
Then, the BSs exhibit a spiraling (or oscillating) pattern of the
frequency dependence of both charge and mass, where
both tend to some finite limiting values at the centers of the spirals.
Qualitatively, the appearance of the
frequency-mass spiral may be related to oscillations in the force-balance
between the repulsive scalar interaction and the gravitational attraction in equilibria.
Indeed,
radially excited rotating BSs do not exhibit a spiraling behavior;
instead the second branch extends all
the way back to the upper critical value of the frequency $\omega_{\rm max}$, forming a loop \cite{Collodel:2017biu}.

\medskip

The main purpose of this paper is to report on the existence of
a new type of solutions, which correspond to {\it chains} of BSs.\footnote{Similar chains, but for a scalar field without self-interactions, were recently studied in~\cite{Herdeiro:2020kvf},  in the context of multipolar BSs. Here, we emphasise the interpretation of multiple BSs, rather than a single multipolar BS.}
These are static, axially symmetric equilibrium configurations
interpreted as a number of BSs  located
symmetrically with respect to the origin, along the symmetry axis.
We construct these solutions and investigate their physical properties
for a choice of the scalar field potential with quartic and sextic self-interaction terms,
which was employed in most of the $Q$-balls literature.
We note that similar configurations of chains of constituents are known to exist
both for gravitating and flat space non-Abelian monopoles and dyons
\cite{Kleihaus:1999sx,Kleihaus:2000hx,Kleihaus:2003nj,Kleihaus:2003xz,Kleihaus:2004is,Teh:2004bq,Paturyan:2004ps,Kleihaus:2004fh,Kleihaus:2005fs,Kunz:2006ex,Kunz:2007jw,Lim:2011ra,Teh:2014zea},
Skyrmions \cite{Krusch:2004uf,Shnir:2009ct,Shnir:2015aba},
electroweak sphalerons \cite{Kleihaus:2008gn,Ibadov:2008hj,Ibadov:2010ei,Ibadov:2010hm,Teh:2014saa},
$SU(2)$ non-self dual configurations \cite{Radu:2006gg,Shnir:2007zz}
and Yang-Mills solitons in ADS$_4$ spacetime
\cite{Kichakova:2012pm,Kichakova:2014fta}.

In these multi-component BSs configurations the constituents
form a chain along the symmetry axis and, consequently,
the scalar field is required to possess  'nodes' (zeros of the scalar field amplitude).
Configurations with $k_z$ nodes on the symmetry axis possess $k_z+1$ constituents. Configurations with even and odd numbers
of constituents can show a different pattern, when their domain of existence
is mapped out.
In particular, we find that the pattern exhibited by a mass-frequency diagram of the
chains of BSs can differ both from the typical spiraling picture and from the closed loop scenario.
For chains with an even number of constituents the
pattern always terminates at critical solutions.
For chains with an odd number of constituents, the pattern depends on
the strength of the gravitational interaction.
The configurations then either merge with the corresponding radial excitation of the fundamental solution,
or the central constituent of the configurations exhibits oscillations while retaining smaller satellite constituents.

\medskip

This paper is organized as follows.
In Section II the theoretical setting is specified.
This includes the action, the equations of motion,
the  Ansatz and the boundary conditions for the BS chains.
The  numerical results for these new equilibrium configurations are shown in Section III .
We give our conclusions in Section IV.

\section{The model}
\subsection{Action, field equations and global charges}

We consider a self-interacting complex scalar field $\Phi$, which is minimally
coupled to Einstein's gravity in a
$(3+1)$-dimensional space-time. The corresponding action of
the system is
\begin{equation}
\label{action}
\mathcal{S}=\int  d^4x \sqrt{-g}\left[ \frac{R}{16\pi G}
   -\frac{1}{2} g^{\mu\nu}\left( \Phi_{, \, \mu}^* \Phi_{, \, \nu} + \Phi _{, \, \nu}^* \Phi _{, \, \mu} \right) - U(|\Phi|^2 )
 \right] ,
\end{equation}
where $R$ is the Ricci scalar curvature, $G$ is Newton's constant,
the asterisk denotes complex conjugation,
and $U $ denotes the scalar field potential.

Variation of the action \re{action} with respect to the metric leads to the Einstein equations
\be
\label{Einstein}
E_{\mu\nu}\equiv R_{\mu\nu}-\frac{1}{2}g_{\mu\nu}R-8 \pi G~T_{\mu\nu}=0 \ ,
\ee
where
\be
\label{SET}
T_{\mu\nu}\equiv
 \Phi_{ , \mu}^*\Phi_{,\nu}
+\Phi_{,\nu}^*\Phi_{,\mu}
-g_{\mu\nu}  \left[ \frac{1}{2} g^{\sigma\tau}
 ( \Phi_{,\sigma}^*\Phi_{,\tau}+
\Phi_{,\tau}^*\Phi_{,\sigma} )+U(|\Phi|^2) \right] \, ,
\ee
is the stress-energy tensor of the scalar field.

The corresponding equation of motion of the scalar field
is the non-linear Klein-Gordon equation
\be
\label{scaleq}
    \left(\Box - \frac{d U}{d |\Phi|^2} \right)\Phi=0\, ,
\ee
where $\Box$ represents the covariant d'Alembert operator.

\medskip

The solutions considered in this work have a static line-element
(with a  timelike Killing vector field  $\xi$),
  being
topologically trivial and
  globally regular, $i.e.$ without an event horizon or conical singularities,
while the scalar field is finite and smooth everywhere.
Also, they approach asymptotically the Minkowski spacetime background.
Their mass $M$
can be obtained from the respective Komar expressions \cite{wald},
\begin{equation}
\label{komar}
{M} = \frac{1}{{4\pi G}} \int_{\Sigma}
 R_{\mu\nu}n^\mu\xi^\nu dV~.
\end{equation}
Here $\Sigma$ denotes a  spacelike hypersurface
(with  the  volume element $dV$),
while
$n^\mu$ is normal to $\Sigma$, with $n_\mu n^\mu = -1$
%
\cite{wald}.
After replacing in (\ref{komar}) the Ricci tensor by the
stress-energy tensor, via the Einstein equations (\ref{Einstein}), one finds
the equivalent expression
\begin{eqnarray}
 \label{komarM2}
&&
M
= \, 2 \int_{\Sigma} \left(  T_{\mu \nu}
-\frac{1}{2} \, g_{\mu\nu} \, T_{\gamma}^{\ \gamma}
 \right) n^{\mu }\xi^{\nu} dV \ .
\end{eqnarray}

The action \re{action} is invariant with respect to the global $\mathrm{U}(1)$
transformations of the complex scalar
field, $\Phi\to\Phi e^{i\alpha }$, where $\alpha$ is a constant.
The following Noether 4-current is associated
with this symmetry
\be
\label{Noether}
j_\mu =- i(\Phi\partial_\mu\Phi^\ast-\Phi^\ast\partial_\mu\Phi)\, .
\ee
It follows that integrating the timelike component of this 4-current in a spacelike slice
$\Sigma$ yields a second conserved quantity -- the \textit{Noether charge}:
\begin{eqnarray}
\label{Q}
Q =\int_{\Sigma} j^{\mu}n_\mu dV.
\end{eqnarray}

\subsection{The ansatz and equations}

Apart from being static, the configurations in this work are also
 axially
symmetric\footnote{The scalar field possesses a time dependence (with $\partial_t \Phi=-i \omega \Phi$),
which disappears at the level of the energy-momentum tensor.
However,  the scalar field is axially symmetric, $\partial_\varphi \Phi=0$.
}.
Thus, in a system of adapted coordinates,
 the space-time  possesses
 two commuting Killing vector fields
\begin{eqnarray}
\xi=\partial/\partial t~~{\rm and }~~\eta=\partial/\partial \varphi ,
\end{eqnarray}
with $t$ and $\varphi$
the time and azimuthal coordinates, respectively.
Line elements
with these symmetries are usually studied
by employing a
metric ansatz \cite{book}
\begin{eqnarray}
\label{metric-i}
ds^2=-e^{-2U(\rho,z)}  dt^2+e^{2U(\rho,z)}\left[e^{2k(\rho,z)}(d\rho^2+dz^2)+P(\rho,z)^2d\varphi^2 \right],
\end{eqnarray}
where $(\rho,z)$ correspond, asymptotically, to the usual cylindrical coordinates\footnote{
 In the Einstein-Maxwell theory, it is always possible
to set $P\equiv \rho$ ,
such that only two independent metric functions appear in the equations,
with $(\rho,z)$  the canonical Weyl coordinates \cite{book}.
For a (complex) scalar field matter content, however,
the generic metric ansatz (\ref{metric-i}) with three independent functions is needed.
}.
However,
it the numerical treatment of the Einstein-Klein-Gordon equations,
it is useful to
employ
 `quasi-isotropic' spherical  coordinates $(r,\theta)$,
defined by the coordinate transformation in (\ref{metric-i})
\begin{eqnarray}
\label{transf}
 \rho=r\sin \theta,~~z=r\cos \theta~,
\end{eqnarray}
with  the usual range $0\leqslant r<\infty$, $0\leqslant \theta \leq \pi$.
Then the metric can then be written in a Lewis-Papapetrou form, with
\be
\label{metrans}
ds^2=-f dt^2 +\frac{m}{f}\left(dr^2+r^2 d\theta^2\right)
+\frac{l}{f}  r^2\sin^2 \theta d\varphi^2~.
\ee
The three metric functions $f$, $l$,  and $m$
are functions of the variables $r$ and $\theta$ only, chosen such that
the trivial angular and radial dependence of the (asymptotically flat)
line element is already factorized.
The relation between the metric functions in the above line-element  and those
in (\ref{metric-i})
is
$f=e^{-2U}$,
$l r^2 \sin^2 \theta=P^2$,
$m=e^{2k}$.
The symmetry axis of the spacetime is located at the set of  points with vanishing norm of $\eta$, $||\eta||=0$;
it corresponds to the $z$-axis in \eqref{metric-i} or the set with $\theta=0,\pi$ in~\eqref{metrans}.
The Minkowski spacetime background is approached for $r\to \infty$, with $f=l=m=1$.

\medskip

For the scalar field we adopt an Ansatz with a harmonic time dependence,
while the amplitude depends  on $(r,\theta)$,
\be
\label{scalans}
\Phi=\phi(r,\theta)e^{-i\omega t},
\ee
where $\omega \geq 0$ is the angular frequency.
The corresponding stress-energy tensor is static,
with the nonvanishing
components
\begin{eqnarray}
\nonumber
&&
T_{r}^r=\frac{f}{m}
\bigg(
\phi_{,r}^2-\frac{\phi_{,\theta}^2}{r^2}
\bigg)
+\frac{\omega^2\phi^2}{f}-U(\phi^2),
~~
T_{\theta}^\theta=\frac{f}{m}
\bigg(
-\phi_{,r}^2+\frac{\phi_{,\theta}^2}{r^2}
\bigg)
+\frac{\omega^2\phi^2}{f}-U(\phi^2),~~
T_{r}^\theta=\frac{2f}{r^2 m}
\phi_{,r} \phi_{,\theta}
\\
\label{Tab}
&&
T_{\varphi}^\varphi=-\frac{f}{m}
\bigg(
\phi_{,r}^2+\frac{\phi_{,\theta}^2}{r^2}
\bigg)
+\frac{\omega^2\phi^2}{f}-U(\phi^2),
~~
T_{t}^t=
-\frac{f}{m}
\bigg(
\phi_{,r}^2+\frac{\phi_{,\theta}^2}{r^2}
\bigg)
-\frac{\omega^2\phi^2}{f}- U(\phi^2).
\end{eqnarray}

After inserting the ansatz (\ref{metrans}), (\ref{scalans})
into the field equations (\ref{Einstein}), (\ref{scaleq})
we find a  system of six coupled partial differential equations
that needs to be solved.
There are three equations for the metric functions $f,l,m$, found by
taking the following suitable combinations of the Einstein equations,
$E_r^r+E_\theta^\theta=0$,
$E_\varphi^\varphi=0$ and
$E_t^t=0$, yielding
\begin{eqnarray}
\label{eqf}
\nonumber
&&
f_{,rr}+\frac{f_{,\theta \theta}}{r^2}
+\frac{2f_{,r}}{r}+\frac{\cot \theta f_{,\theta}}{r^2}
-\frac{1}{f} \bigg(f_{,r}^2+\frac{f_{,\theta}^2}{r^2} \bigg)
+\frac{1}{2l} \bigg(f_{,r}l_{,r}+\frac{f_{,\theta}l_{,\theta}}{r^2} \bigg)
+16\pi G   \left(U(\phi^2)-\frac{2\omega^2 \phi^2}{f} \right) m=0~,
\\
\label{eql}
&&
l_{,rr}+\frac{l_{,\theta \theta}}{r^2}
+\frac{3l_{,r}}{r}+\frac{2\cot \theta l_{,\theta}}{r^2}
-\frac{1}{2l} \bigg(l_{,r}^2+\frac{l_{,\theta}^2}{r^2}\bigg)
+32\pi G   \left(U(\phi^2)-\frac{ \omega^2 \phi^2}{f} \right)\frac{l m}{f}=0~,
\\
\label{eqm}
\nonumber
&&
m_{,rr}+\frac{m_{,\theta \theta}}{r^2}
+\frac{m_{,r}}{r}
+\frac{m}{2f^2} \bigg(f_{,r}^2+\frac{f_{,\theta}^2}{r^2}\bigg)
-\frac{1}{m} \bigg(m_{,r}^2+\frac{m_{,\theta}^2}{r^2}\bigg) \\
&&
\ \ \ \ \ \ \ +16\pi G
      \left[ \frac{f}{m}\bigg(\phi_{,r}^2+\frac{\phi_{,\theta}^2}{r^2}\bigg) + U(\phi^2)-\frac{\omega^2 \phi^2}{f} \right]\frac{m^2}{f}=0~,
\nonumber
\end{eqnarray}
and one equation  for the scalar field amplitude
\begin{eqnarray}
\phi_{,rr}+\frac{\phi_{,\theta \theta}}{r^2}
+ \left(\frac{2}{r}+\frac{ l_{,r}}{2l} \right)\phi_{,r}
+ \left(\cot \theta+\frac{ l_{,\theta}}{2l}  \right)  \frac{\phi_{,\theta}}{r^2}
+ (\frac{\omega^2}{f}-\frac{\partial U }{\partial \phi^2})\frac{m}{f}\phi=0 \ .
\end{eqnarray}
Apart from these, there are two more Einstein equations,
  $E_\theta^r =0,~E_r^r-E_\theta^\theta  =0$,
  which are not solved in practice,
	being treated  as constraints and used to check the numerical accuracy of the solutions.

The mass $M$ is computed from the
Komar expression (\ref{komar}),  where we insert the metric ansatz (\ref{metrans}),
with unit vector $n = -\sqrt{f}dt$,
the volume element
$dV =1/ \sqrt{f} \, \sqrt{-g} \, dr \, d\t \, d\varphi$,
and
$\sqrt{-g}=r^2 \sin \theta \frac{\sqrt{l}m}{f}$.
In evaluating (\ref{komar}),
we  use  the fact that
 $R_t^t$  is a total derivative:
\begin{eqnarray}
\nonumber
\sqrt{-g}R_t^t
&=&
-\frac{\partial}{\partial r}
\left(
\frac{ r^2 \sin \theta \sqrt{l} f_{,r}}{2f}
\right)
-\frac{\partial}{\partial \theta}
\left(
\frac{ \sin \theta \sqrt{l} f_{,\theta}}{2f}
\right)
.
\end{eqnarray}
Then it follows that  $M$
can be read off from the asymptotic expansion of the metric function  $f$
\begin{eqnarray}
f = 1- \frac{2MG}{r} + \mathcal{O}\left( \frac{1}{r^2} \right) \ . \ \ \
\label{MQasym1}
\end{eqnarray}
Alternatively, the mass $M$
can be obtained by direct integration of
(\ref{komarM2}) ,
\begin{eqnarray}
 \label{tolman}
M =\int
\left(  T_t^t -T_{r}^{r} -T_{\theta}^{\theta}-T_{\varphi}^{\varphi}
 \right) \, \sqrt{-g}
\, dr \, d\t \, d \varphi
=
4\pi \int_0^\infty dr \int_0^\pi d\theta~
r^2 \sin \theta \frac{\sqrt{l}m}{f}
\left(U(\phi^2)-\frac{2 \omega^2\phi^2}{f} \right)~.
\end{eqnarray}

In terms of the above Ansatz the Noether charge $Q$ is given by
\be
Q=4\pi \omega \int\limits_0^\infty dr \int\limits_0^\pi d\theta ~\frac{\sqrt{l} m}{f^2} r^2 \sin \theta ~\phi^2 ~.
\ee

Also, let us note that the solutions in this work have no horizon. Therefore they are
zero entropy objects, without an intrinsic temperature.
 Still, in analogy to black holes, one may write a ``first
law of thermodynamics" \cite{Lee:1991ax}, albeit without the entropy term, which reads
\begin{equation}
\label{first-law}
dM = \omega dQ ~.
\end{equation}
This gives a relation between the mass and Noether  charge of neighbouring  BS solutions which can be used to check the numerical accuracy of the solutions.

\subsection{The potential and scaling properties }

The solutions in this work were found for a potential
originally proposed in \cite{Deppert:1979au,Mielke:1980sa} ,
\be
\label{pot}
U(|\Phi|^2)= \nu |\Phi|^6 -\lambda  |\Phi|^4 + \mu^2  |\Phi|^2
\ee
which  is chosen such that nontopological soliton solutions -  $Q$-balls - exist in the absence of the Einstein term in the action (\ref{action}), $i.e.$ on a Minkowski spacetime background
\cite{Friedberg:1986tq,Volkov:2002aj,Kleihaus:2005me,Kleihaus:2007vk}.
Also,
at least in the spherically symmetric case,
this choice of the potential follows for the existence of very massive and highly compact
objects approaching the black hole limit \cite{Friedberg:1986tq}.

The parameter $\mu$ in (\ref{pot}) corresponds to
 the mass of the scalar excitations around the $|\Phi|=0$ vacuum.
Apart from that, the potential possesses two more parameters $(\nu,\lambda)>0$, determining the self-interactions,
which are chosen
in such a manner that it possesses a local minimum at some finite non-zero
value of the field $|\Phi|$, besides the global minimum at $|\Phi|=0$.

Two of the constants in (\ref{pot})
 can actually be absorbed into a redefinition of the radial coordinate
together with a rescaling of the scalar field,
\begin{eqnarray}
\label{scaling}
r\to \frac{u_0}{\mu} r\ ,~~~
\phi \to \frac{\sqrt{\mu}}{\nu^{1/4} \sqrt{u_0}} \phi \ ,
\end{eqnarray}
with $u_0>0$ an arbitrary constant.
Note that the scalar field frequency changes accordingly,
$
\omega \to \frac{u_0}{\mu} \omega.
$

Then, the potential  (\ref{pot}) becomes (up  to an overall factor),
\begin{eqnarray}
\label{potn}
U =\phi^6 -\bar \lambda \phi^4+ u_0^2 \phi^2,
\qquad  {\rm with}
~~
\bar \lambda=\frac{\lambda u_0}{\mu \sqrt{\nu}} \ .
\end{eqnarray}
The  choice employed in most of the
$Q$-ball literature
 is
\begin{eqnarray}
\label{choice}
 u_0^2=1.1\ ,~~{\rm and}~~
 \bar \lambda=2 \ ,
\end{eqnarray}
which are the values
 used also in this work.

For completeness, let us mention that
for the specific ansatz  (\ref{metrans}), (\ref{scalans})  with the above scalings,
the equations for gravitating $Q$-balls (which are effectively solved)
can   also  be derived by extremizing the following reduced action\footnote{In (\ref{Leff-gs})  we use the dimensionless radial variable $r$
and the dimensionless frequency $\omega$ given in units set by $\mu$.
 }
\begin{eqnarray}
\label{Leff}
S_{red}=\int dr d \theta \left({\cal L}_g-4 \alpha^2 {\cal L}_s \right),
\end{eqnarray}
with [for ease of notation, we denote  $(\nabla S)\cdot (\nabla T)\equiv
S_{,r}T_{,r}+\frac{1}{r^2}S_{,\theta}T_{,\theta}$]:
\begin{eqnarray}
\nonumber
{\cal L}_g&=&r^2 \sin \theta \sqrt{l}
 \bigg [
\frac{1}{2 lm}(\nabla l)\cdot (\nabla m)
-\frac{1}{2f^2}(\nabla f)^2
\\
\label{Leff-gs}
&&{~~~~~~~~~~~~}
-\frac{1}{r l}\left(l_{,r}+\frac{\cot \theta}{r}l_{,\theta}\right)
+\frac{1}{r m}\left(m_{,r}+\frac{\cot \theta}{r}m_{,\theta}\right)
 \bigg ],
\\
\nonumber
{\cal L}_s&=& r^2 \sin \theta\frac{m \sqrt{l}}{f}
\bigg [
\frac{f}{m}   (\nabla \phi)^2
 -\frac{\omega^2 \phi^2}{f}+\phi^6-2\phi^4+1.1 \phi^2
\bigg].
\end{eqnarray}
The   (dimensionless) coupling
constant reads
\begin{eqnarray}
\alpha^2=\frac{4\pi G \mu}{\sqrt{\nu} u_0},
\end{eqnarray}
which determines the strength of  the gravitational coupling of the solutions.

As with the known spherically symmetric configurations, solutions exist for $0\leqslant \alpha<\infty$.
The limit $\alpha \to 0$
corresponds to the non-backreacting case,
$i.e.$ $Q$-balls on a fixed Minkowski background.
To understand the large $\alpha$ limit,
we define
 $\hat \phi =\phi/\alpha $.
Then
for large values of the
effective gravitational coupling $\alpha$, the  non-linearity of the potential \re{pot} becomes suppressed
and the system approaches
 the usual Einstein-Klein-Gordon model \cite{Kaup:1968zz,Ruffini:1969qy}, with its corresponding mBS solutions.
That is, the resulting  equations
are identical to those found for a model with a non-self-interacting, massive complex scalar field $\hat \phi$.
However, in this work we shall restrict our study to the case of a finite, nonzero $\alpha$.
The basic properties of the mBS chains were studied (in a more general context) in the recent work
\cite{Herdeiro:2020kvf},
while   the issue of flat spacetime $Q$-ball chains will be discussed elsewhere.

\subsection{Boundary conditions }

The solutions studied in this work
are globally regular and asymptotically flat,  possessing finite mass  and
Noether charge.
Appropriate boundary conditions guarantee that these conditions are satisfied.

Starting with the boundary conditions for the metric functions,
regularity of the solutions at the origin requires
\be
\label{bcor}
 \partial_r f\bigl.\bigr|_{r=0}=\partial_r m\bigl.\bigr|_{r=0}
=\partial_r l\bigl.\bigr|_{r=0} = 0\, .
\ee
Demanding asymptotic flatness at spatial infinity yields
\be
\label{bcinf}
f\bigl.\bigr|_{r\to \infty}=m\bigl.\bigr|_{r\to \infty}
=l\bigl.\bigr|_{r\to \infty}=1.
\ee
Axial symmetry and regularity impose
on the symmetry axis at $\theta=0,\pi$ the conditions
\be
\label{bcpole}
\partial_\theta \phi\bigl.\bigr|_{\theta = 0,\pi} =
\partial_\theta f\bigl.\bigr|_{\theta = 0,\pi} =
\partial_\theta m\bigl.\bigr|_{\theta = 0,\pi} =
\partial_\theta l\bigl.\bigr|_{\theta = 0,\pi} =0 \, .
\ee
Further, the condition of the absence of a conical singularity
requires that the solutions should satisfy the constraint
$m\bigl.\bigr|_{\theta = 0,\pi}=l\bigl.\bigr|_{\theta = 0,\pi}$.
In our numerical scheme we explicitly verified (within the numerical accuracy) this condition
on the symmetry axis.

Turning now to the  boundary conditions for the scalar field amplitude,
we mention first that $\phi$ approaches  asymptotically the global minimum,
\be
\label{bcinf1}
\phi \bigl.\bigr|_{r\to \infty}=0~,
\ee
while on the symmetry axis we impose
\be
\label{bcz}
\partial_\theta \phi \bigl.\bigr|_{\theta=0,\pi}=0 \  .
\ee
The behaviour of the scalar field at the origin is more complicated,
depending on the considered parity ${\cal P}$.
As mentioned in the Introduction,
the solutions   split into two classes, distinguished by their behaviour
$w.r.t.$ a reflection along the equatorial plane $\theta=\pi/2$.
The geometry is left invariant under this transformation,
 \begin{eqnarray}
\label{parity0}
f (r, \pi-\t) = f (r, \t),~~l(r, \pi-\t) = l(r, \t) ,~~m(r, \pi-\t) = m(r, \t)  ~;
\end{eqnarray}
 for the scalar field, however, there are two possibilities:
\begin{eqnarray}
\label{parity}
&&
{\cal P}=1~~~{\rm (even~parity)}:~~
\f (r, \pi-\t) = {\phantom{-}} \f (r, \t) \ ,
\\
&&
{\cal P}=-1~{\rm (odd~parity)}:~~
\f (r, \pi-\t) =          -   \f (r, \t)
~.
\end{eqnarray}
We use this symmetry to reduce the domain of integration to $ [0,\infty)\times [0,\pi/2]$,
with
\be
\label{bcplane}
\partial_\theta f\bigl.\bigr|_{\theta = \pi/2} =
\partial_\theta m\bigl.\bigr|_{\theta = \pi/2} =
\partial_\theta l\bigl.\bigr|_{\theta = \pi/2} =0 \, ,
\ee
and
\be
\label{k-gen}
{\cal P}=1:~\partial_\theta \phi \bigl.\bigr|_{\theta = \pi/2} =0,~~~
{\cal P}=-1:~ \phi \bigl.\bigr|_{\theta = \pi/2} =0,~~
\ee
%
together with
\begin{eqnarray}
\label{k-even}
{\cal P}=1:~
\partial_r\phi \bigl.\bigr|_{r=0}=0~ \ ,
~~~
{\cal P}=-1:~ \phi \bigl.\bigr|_{r=0}= 0 \, .
\end{eqnarray}
Finally, let us mention that,
for both ${\cal P}=\pm 1$,
one can formally construct an
 approximate expression of the solutions
compatible with the above boundary conditions, $e.g.$ by assuming the existence of a power series form close to $r=0$.
The resulting expressions
are of little help in understanding the properties of the solutions,
and a numerical approach is necessary\footnote{We recall that,
more than fifty years after their discovery \cite{Kaup:1968zz,Ruffini:1969qy},
the static mBS are still not known in closed form.
 }.
However, one important result of this study is the bound-state condition
\begin{eqnarray}
 \omega\leqslant \mu,
\end{eqnarray}
which emerges from the finite mass requirement.
No similar result is found for the minimal value of frequency.

\subsection{Numerical method}

The set of four coupled non-linear
elliptic partial differential equations
for the functions $(f,l,m;\phi)$
has been solved numerically
subject to the boundary conditions defined above.
In a first stage, a new compactified radial variable $x$ is introduced, replacing $r$,  with
\begin{eqnarray}
x\equiv \frac{r}{c+r},
\end{eqnarray}
with $c$ an arbitrary parameter, typically of order one.
With this choice, the
semi-infinite region $[0,\infty)$ is mapped to the finite interval $[0,1]$.
This avoids the use of a cutoff radius.

The numerical calculations have been
performed by using a using
a sixth-order finite difference scheme.
The system of equations is discretized on a grid
with a typical number of points $129\times 89$.
The underlying linear system is solved with the Intel MKL
PARDISO sparse direct solver \cite{pardiso} using the Newton-Raphson method. Calculations are performed with
the packages FIDISOL/CADSOL \cite{schoen} and CESDSOL\footnote{Complex Equations -- Simple
Domain partial differential equations SOLver is a C++ package
being developed by one of us (I.P.).
} library.
In all cases, the typical
errors are of order of $10^{-4}$.

For the choice (\ref{choice})
of the potential's parameters,
the only input parameters are
the gravitational coupling constant
$\alpha$
together with the scalar field's frequency
$\omega$.
The parity of the solutions is
imposed via the boundary conditions (\ref{parity}), (\ref{k-even}).
The number of individual
constituents
results from the numerical output.
Also, in all plots below, quantities are given in natural units set by $\mu,G$.

\section{The solutions}

\subsection{Nodal structure and energy distribution}

The choice of the parity
${\cal P}$
is related to the number of distinct constituents of the solutions
(as resulting $e.g.$ from the spatial distribution of the energy density).
Moreover,
denoting the number of nodes on the symmetry axis by $k_z$,
the  solutions can be classified by the number of nodes $k_z$.
The number of constituents of the chains
is given by $k_z+1$.
The even-parity configurations (${\cal P}=1$)
have an even $k_z$,
while the solutions with an  odd $k_z$
are found for
${\cal P}=-1$.
For example,
 the spherically symmetric solutions have $k_z=0$
and one single constituent localized at the origin, $r=0$.
The simplest non-spherical configuration has $k_z=1$ and
represents a pair of static BSs\footnote{
In principle, the $k_z=1$ solutions can be  thought of as the
static limit of
negative parity spinning configurations considered in \cite{Kleihaus:2007vk}.
} with opposite phases, the inversion of the sign of the scalar field function $\phi$ under reflections $\theta \to \pi - \theta$ corresponds to the shift of the phase $\omega t \to \omega t +\pi$.  

It was pointed out that the character of the interaction between Q-balls
in Minkowski spacetime depends on their relative phase
\cite{Battye:2000qj,Bowcock:2008dn}.  If the Q-balls are in phase, the interaction is attractive,
if they are out of phase,
there is a repulsive force between them.
Thus, an axially symmetric $k_z=1$ 
 solution can be balanced by the gravitational attraction.

Solutions with $k_z>1$ exist as well; the maximal value we have reached
so far is $k_z=5$;
 however,  they are likely to exist for an arbitrarily large $k_z$.

 \medskip

To illustrate these aspects,
we  display in Fig.~\ref{functions-top}
several functions of interest for
the   five types of representative configurations.
Both odd and even parity configurations
are shown there, with
the node numbers $k_z=0 - 4$;
also the solutions have
  the same values of the input parameters,
	$\alpha=0.25$ and $\omega/\mu=0.8$, being located on the 1st branch in the $(\omega, M)$-diagram (as described below).
%
These chains possess one to five constituents (from top to bottom),
as seen by the number of peaks of the charge density, shown
in the left panels.
The middle panels represent the scalar field amplitude $\phi$,
and the right panels show the metric function $f=-g_{tt}$.
For the sake of clarity we have chosen to exhibit these figures
in polar coordinates $(\rho,z)$, as given by eq. \eqref{transf}.

\begin{figure}[t!]
    \begin{center} 
   \includegraphics[height=15.cm]{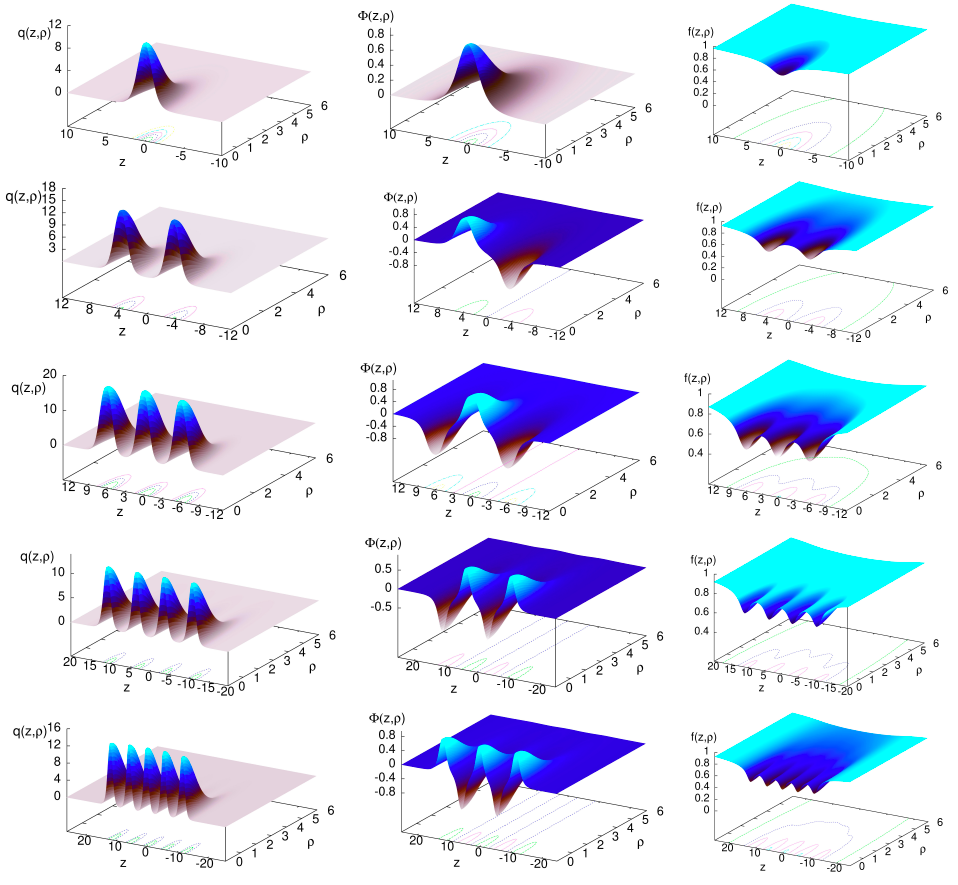}
    \end{center}
    \caption{\small
Chains of BSs with one to five constituents (from top to bottom)
on the  first (a.k.a. \textit{fundamental}) branch for $\alpha=0.25$ at $\omega/\mu=0.80$:
$3d$ plots of the $U(1)$ scalar charge distributions (left plots),
the scalar field functions $\phi$ (middle plots) and the metric functions $f$
(right plots) versus the coordinates $\rho = r\sin \theta$ and $z=r\cos\theta$.}
    \lbfig{functions-top}
\end{figure}

The first row shows a single spherically symmetric BS for comparison.
The second row exhibits the pair of BSs.
The charge density has two symmetric peaks,
the metric function has two symmetric troughs, while the scalar field function
is anti-symmetric, featuring a peak and a trough.
The triplet, quartet and quintet in the next few rows
feature $k_z+1$ very similar (in size and shape) peaks
for the charge density and troughs for the metric function, while
the scalar field shows alternating peaks and troughs,
all located symmetrically along the $z$-axis.
Thus  on the fundamental branch we basically encounter
a chain consisting of $k_z+1$ BSs, all possessing
similar size, shape and distance from their next neighbors.

 \medskip

This  picture partially changes as we move along the domain of solutions, for a given coupling $\alpha$.
This is seen in Fig.~\ref{functions-bottom}, where these chains ($k_z=0 - 4$) are now
shown 
for illustrative solutions sitting on the \textit{second} branch of the $M$ $vs.$ $\omega$ domain  of  existence  (see the  discussion in the next subsection),  with $\alpha=0.25$ for $k_z=0-3$, $\alpha=0.5$ for $k_z=4$,
and $\omega/\mu=\{ 0.43, 0.47, 0.57, 0.7, 0.7\}$ for $k_z=0 - 4$, respectively.
As we look at the charge density of the configurations, we see a dominant
peak at the center for odd chains and a dominant inner pair for even chains,
while the other peaks of the triplet, quartet and quintet have turned into slight
elevations, hardly visible in the figures.
In the metric functions the troughs at the center of the odd chains dominate,
and likewise the (almost merged) inner pair of troughs of the even chains.
All other troughs are weakened substantially.
The scalar field itself, however, retains the outer peaks and troughs to
a somewhat greater extent,  still reflecting clearly the number of constituents
of the chains.

\begin{figure}[t!]
    \begin{center} 
   \includegraphics[height=15.cm]{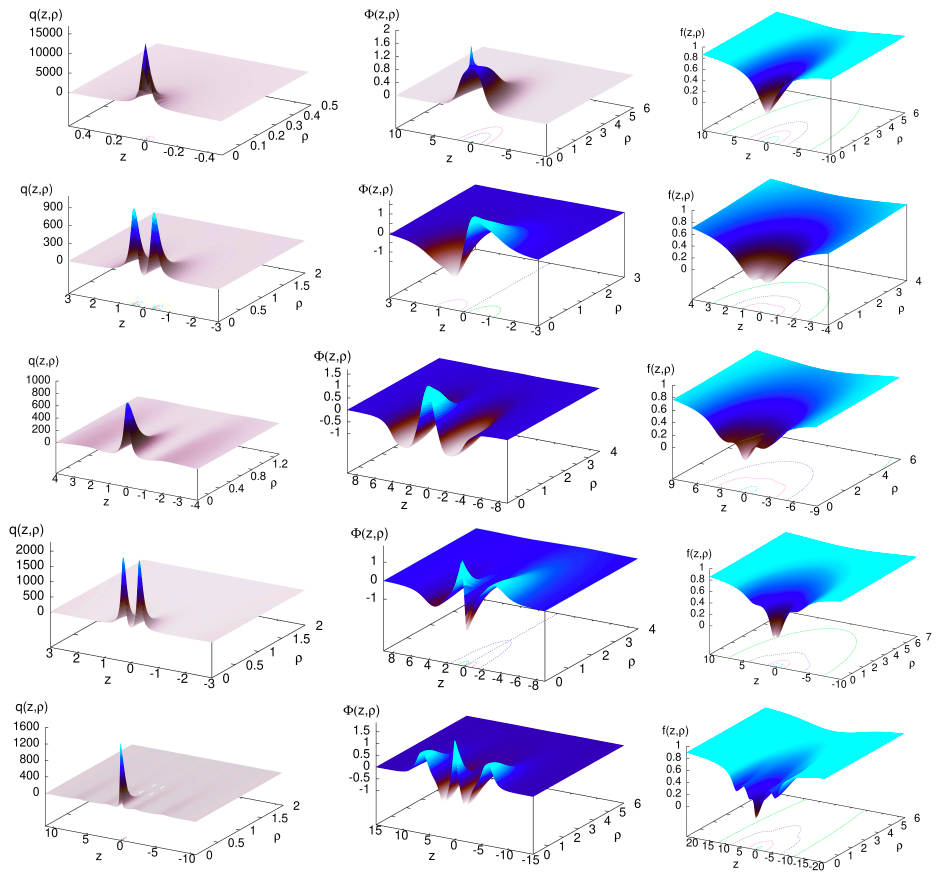}
    \end{center}
    \caption{\small
Chains of BSs with one to five constituents (from top to bottom)
on the second branch for
differing values of $\alpha$ and $\omega$:
$3d$ plots of the $U(1)$ scalar charge distributions (left plots),
the scalar field functions $\phi$ (middle plots) and the metric functions $f$
(right plots) versus the coordinates $\rho = r\sin \theta$ and $z=r\cos\theta$.}
    \lbfig{functions-bottom}
\end{figure}

\subsection{The $\omega$-dependence and the  branch structure}

Recall the frequency dependence for the single BSs and
for fixed coupling $\alpha$ \cite{Friedberg:1986tq}.
The set of BSs emerges from the vacuum at the maximal frequency,
given by the boson mass
$\omega_\text{max}=\mu$.
Thus, unlike the case of $Q$-balls in flat space, where mass and charge diverge,
these quantitites vanish in this limit.
Decreasing  the frequency $\omega$ spans the first or fundamental branch, which terminates at the first backbending of the  curve, at which point it moves towards larger frequencies. The curve then follows a spiraling/oscillating pattern, with successive backbendings.

The mass and charge form a spiral, as $\omega$ is varied,
while the minimum of the metric function $f$
and the maximum of the scalar function $\phi$ shows damped oscillations.
The set of solutions tends to a limiting solution at the center of the spiral
which has finite values of the mass and charge.
However, the values of the scalar field function $\phi$ and the metric function $f$ at the center of the star,
which represent the maximal and minimal values of these functions,
$\phi_\text{max}$ and $f_\text{min}$, respectively,
do not seem to be finite, with
$\phi_\text{max}$ diverging and $f_\text{min}$ vanishing in the limit.

Let  us now consider  the frequency dependence of the BS  chains,
when the coupling $\alpha$ is kept fixed.
Like the single BSs,
all chains emerge similarly from the vacuum at the maximal frequency, given by the boson mass
$\omega_\text{max}=\mu$,
where their mass and charge vanish in the limit.

When $\omega$ is decreased, mass and charge rise and the chains follow along their fundamental branch.
As for the single BS,
for all chains this fundamental branch ends at a minimal value of the frequency,
from where a second branch arises.
But then even and odd chains will in general exhibit different patterns, which will also depend  on the coupling strength $\alpha$.

\subsubsection{Even chains}


\begin{figure}[t!]
    \begin{center} 
        \includegraphics[height=.35\textwidth, trim = 20 20 100 20, clip = true]{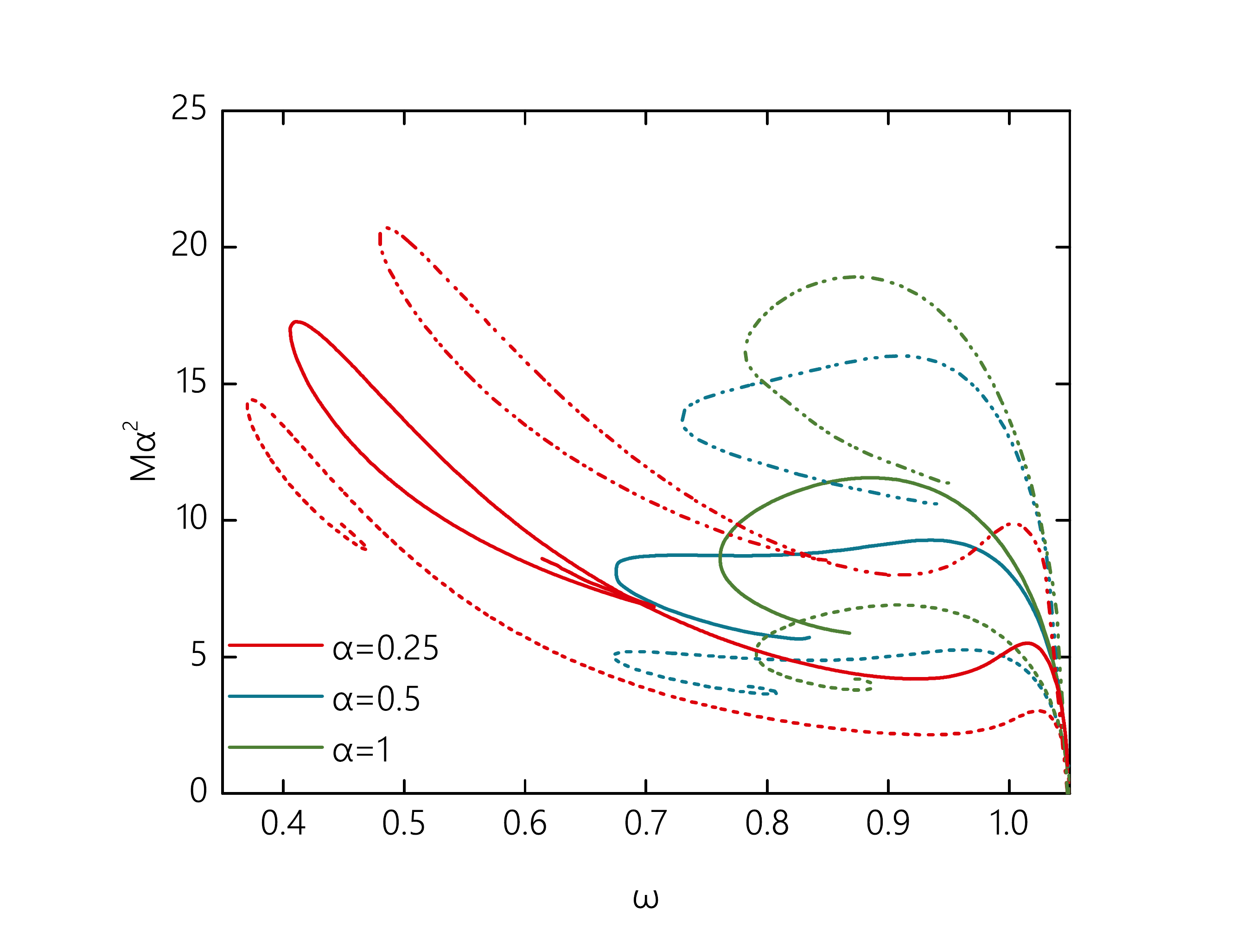}
        \includegraphics[height=.35\textwidth, trim = 20 20 100 20, clip = true]{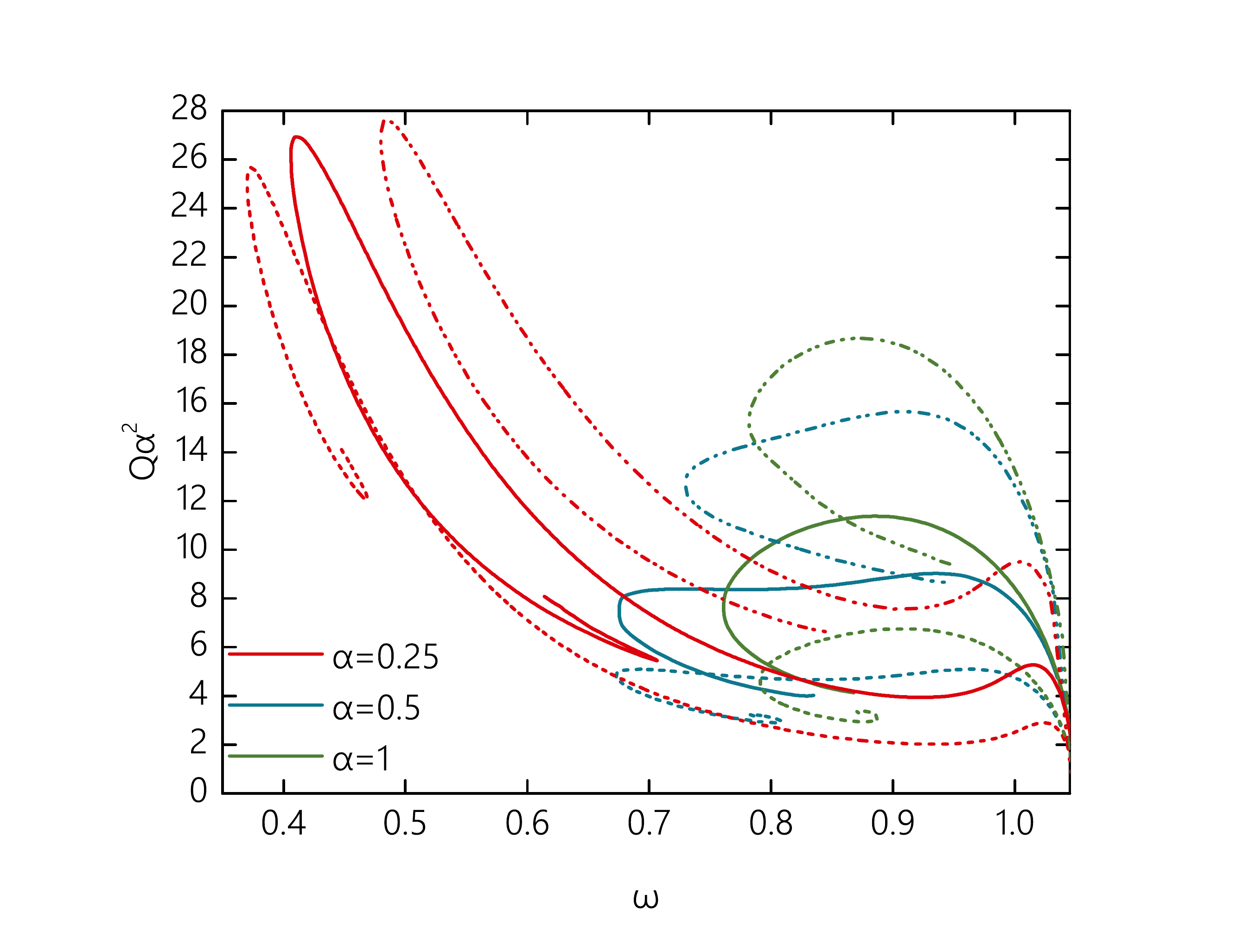}
        \includegraphics[height=.35\textwidth, trim = 20 20 100 20, clip = true]{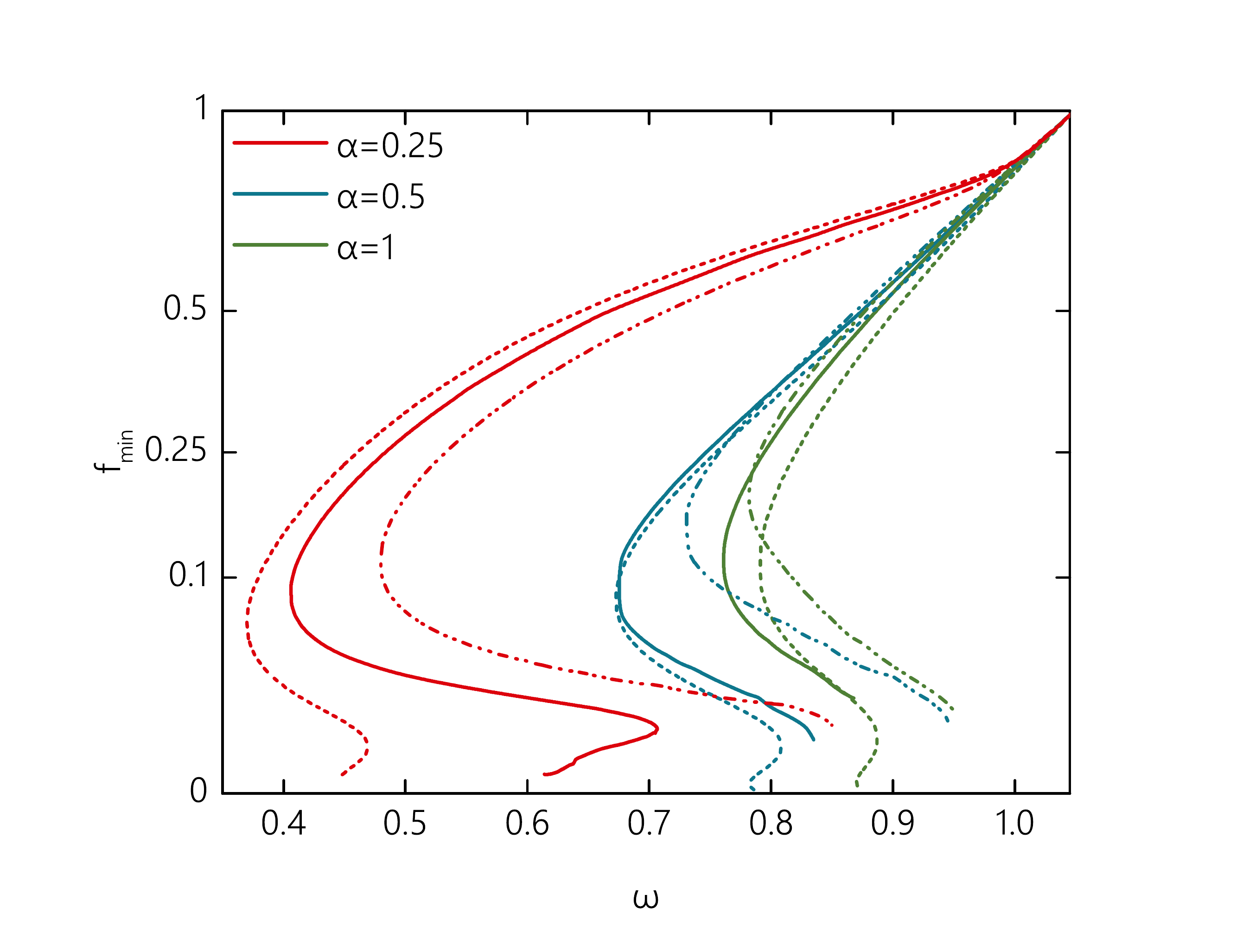}
        \includegraphics[height=.35\textwidth, trim = 20 20 100 20, clip = true]{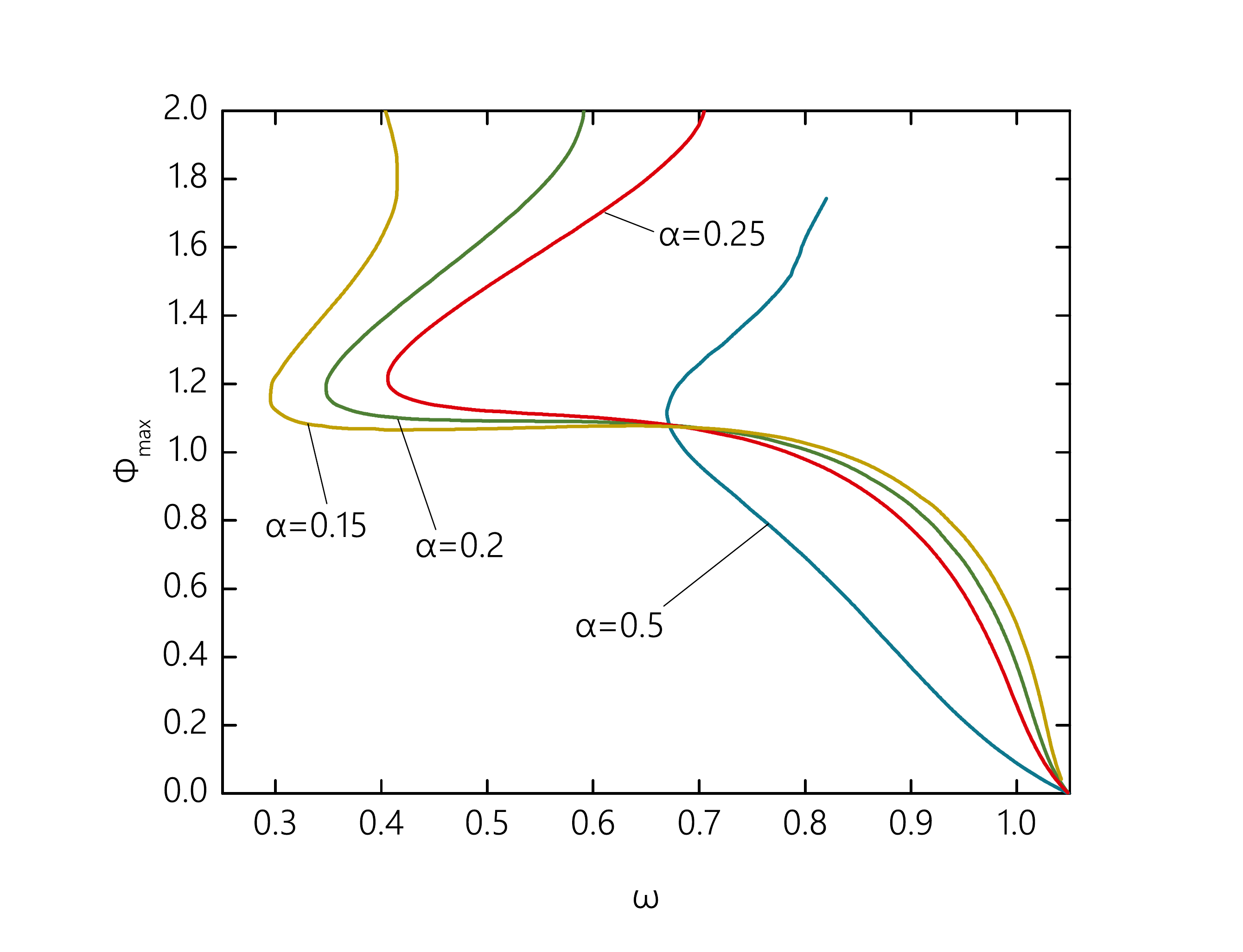}
        \includegraphics[height=.35\textwidth, trim = 20 20 100 20, clip = true]{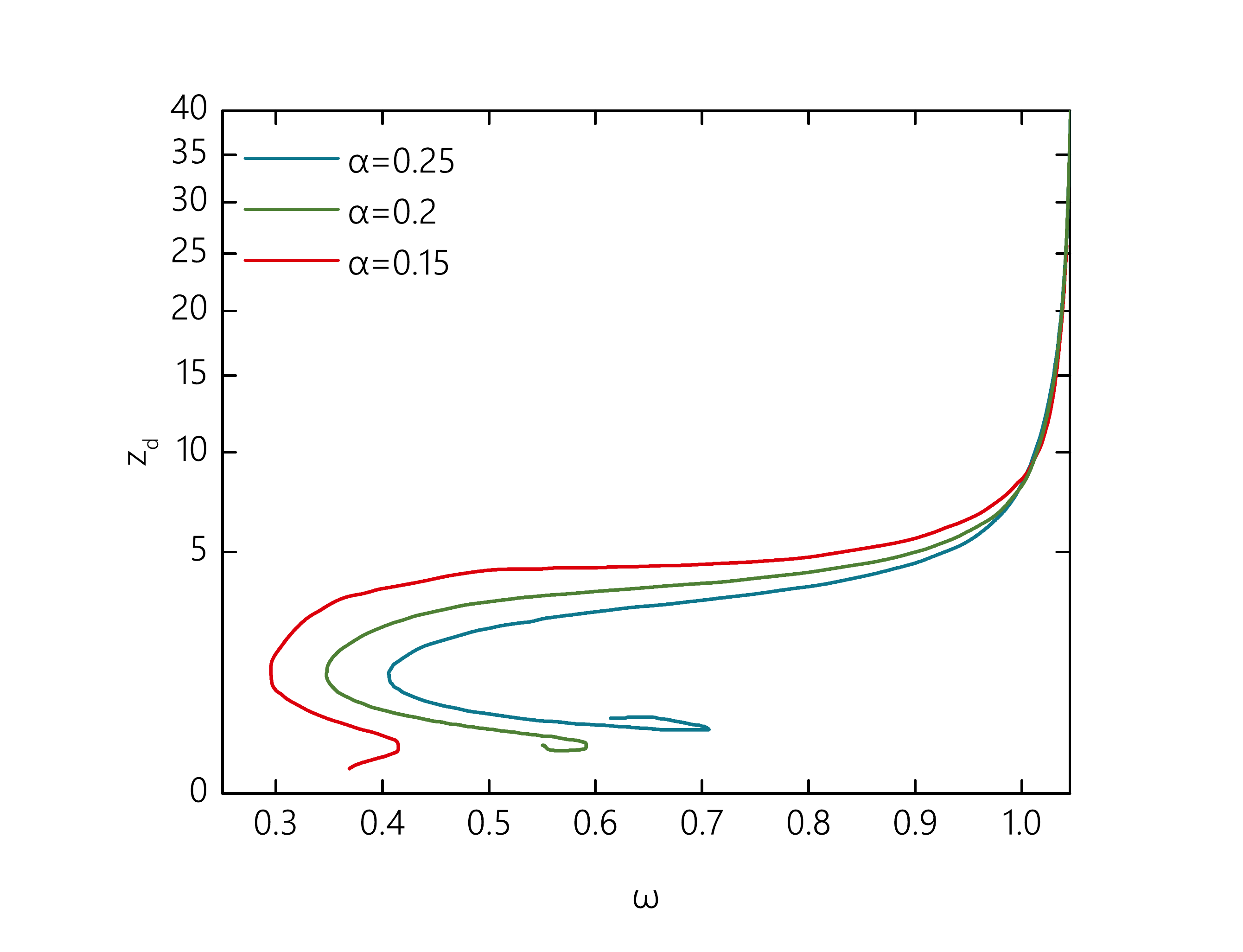}
    \end{center}
    \caption{\small
Comparison of the $k_z=0$ single BSs (dashed curves), the $k_z=1$ pair of BSs (solid curves),
and the $k_z=3$ quartet of BSs (dash-dotted curves):
scaled ADM mass $M$ (upper left panel),
scaled charge $Q$ (upper right panel),
minimal value of the metric function $f_\text{min}$ (middle left panel),
maximal value of the scalar field $\phi_\text{max}$  of the $k_z=1$ pair only (middle right panel),
and separation $z_\text{d}$ between the two components of the $k_z=1$ pair only (lower panel)
$vs.$ frequency $\omega$ for
several values of the coupling $\alpha$.
Note the quadratic scale for $f_\text{min}$ and $z_\text{d}$.}
    \lbfig{comparison-2}
\end{figure}

Let us consider the BS pair and the higher even chains in more detail.
We illustrate the $\omega$-dependence for even chains in Fig.~\ref{comparison-2},
selecting the BS pair ($k_z=1$) and the BS quartet ($k_z=3$),
and comparing with the single BS.
In the upper panels we show
the $k_z=1$ pair (solid curves), the $k_z=3$ quartet  (dash-dotted curves),
as well as the $k_z=0$ single BS (dashed curves).
For the latter we only show the first few branches.
In the two upper panels we exhibit the scaled ADM mass $M$ (left)
and the scaled charge $Q$ (right).
The different colors refer to different values of the coupling $\alpha$.
The middle left panel shows in an analogous manner
the minimal value $f_\text{min}$ of the metric function $f$.
Restricting to the $k_z=1$ pair, we then exhibit
the maximal value $\phi_{\rm  max}$ of the scalar field function $\phi$
on right  middle  panel,
and the separation $z_\text{d}$ between the components of the $k_z=1$ pair in the lower panel.
All quantities are shown versus the  frequency $\omega$.

Whereas the single BSs constitute an infinite set of branches,
that form a spiral or a damped oscillation,
depending on the quantity of interest \cite{Friedberg:1986tq},
the even chains seem to end in a limiting configuration
quite abruptly somewhere in the middle of a branch.
The number of branches before this limiting configuration
is encountered depends on the strength of the gravitational coupling $\alpha$.
For small $\alpha$ there are more branches, for larger $\alpha$,
the limiting configuration is already encountered on the second branch,
as illustrated in Fig.~\ref{comparison-2} for the BS pair.

The mass and the charge exhibit only the onset of a spiraling behavior with
two branches for the larger $\alpha$ values
($\alpha=0.5, 1$), and three branches for the smaller ones
($\alpha=0.15, 0.2, 0.25$).
While the minimum $f_\text{min}$ the metric function (middle left)
and the maximum $\phi_\text{max}$ of the scalar function (middle right)
exhibit only two or three oscillations.
The coordinate distance $z_\text{d}$ between the two components of the pair,
as given by twice the value of the $z$-coordinate of the maximum
$\phi_\text{max}$,
exhibits both types of behavior (lower).
For small $\alpha$ ($\alpha=0.15$) $z_\text{d}$ shows three oscillations,
while it decreases continuously. For larger $\alpha$ ($\alpha=0.25$)
it exhibits the onset of a spiral, with again larger values on the third branch.

\medskip

Let us now address the limiting behavior of the pairs in more detail,
and its dependence on the coupling $\alpha$.
To that end, we illustrate in Fig.~\ref{profiles} the profiles of
the metric function $f$ (left panels) and the scalar field function $\phi$ (right panels)
on the $z$-axis, choosing a large (upper panels) and a small (lower panels)
value of the coupling $\alpha$.
Since for the large $\alpha$, the coordinate distance $z_\text{d}$
does not decrease monotonically, but increases again towards the limiting solution,
we retain two well-separated components in the limit. The minimum of the
metric function $f_\text{min}$ tends to zero in the limit,
while the maximum (minimum) of the scalar field function $\phi_\text{max}$
($\phi_\text{min}$) becomes extremely sharp.


\begin{figure}[h!]
    \begin{center}
        \includegraphics[height=.26\textheight]{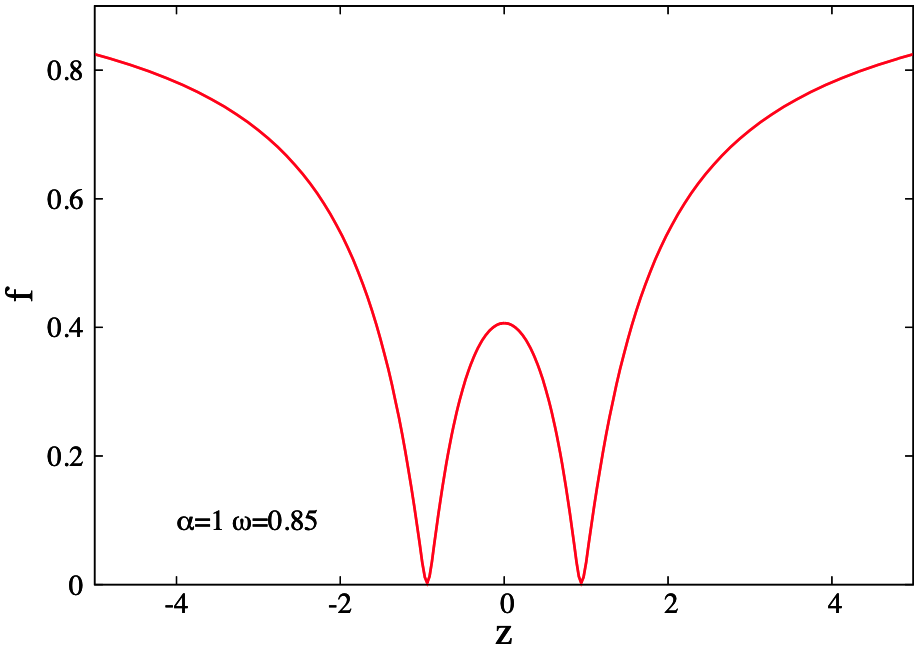}
        \includegraphics[height=.26\textheight]{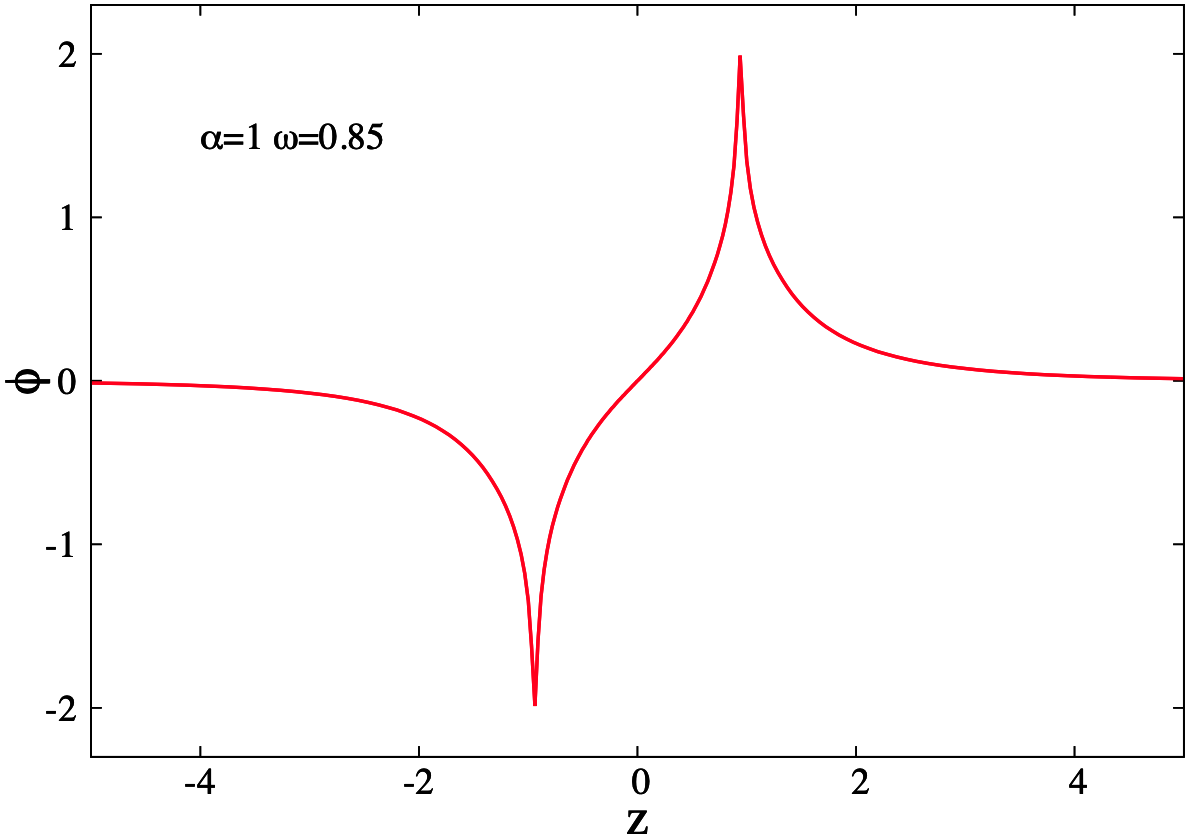}\\
 \includegraphics[height=.26\textheight]{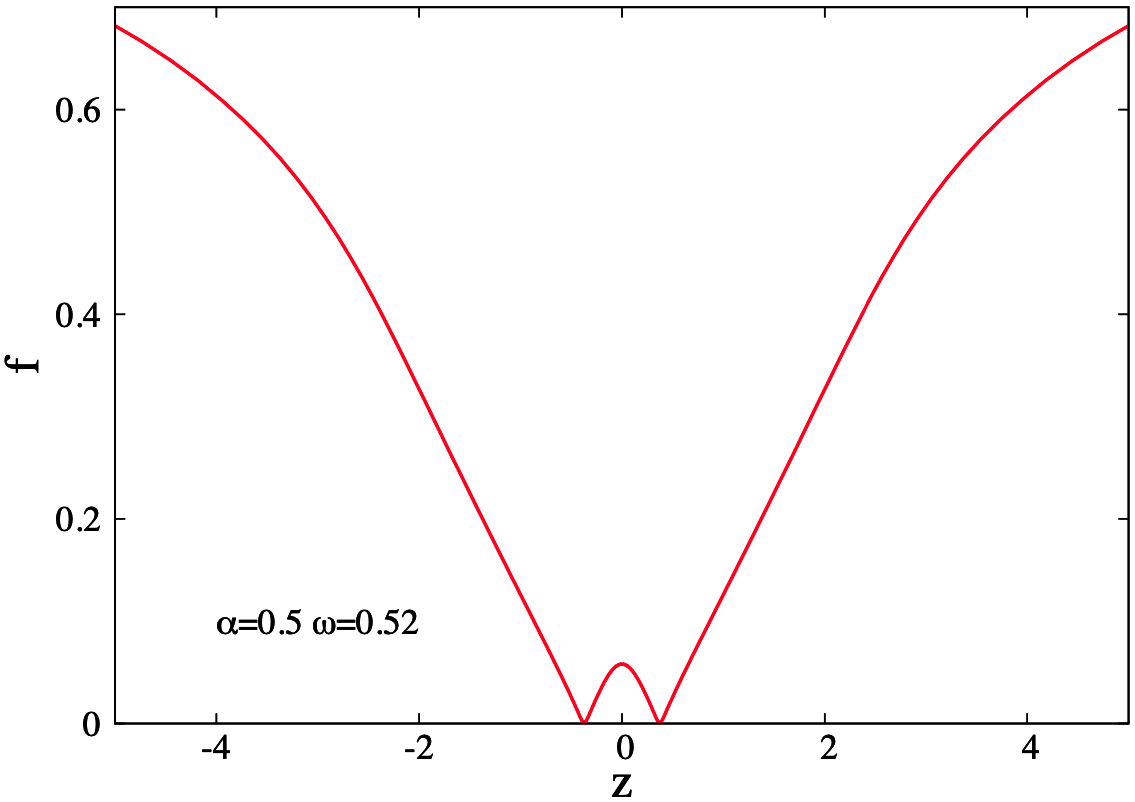}
        \includegraphics[height=.26\textheight]{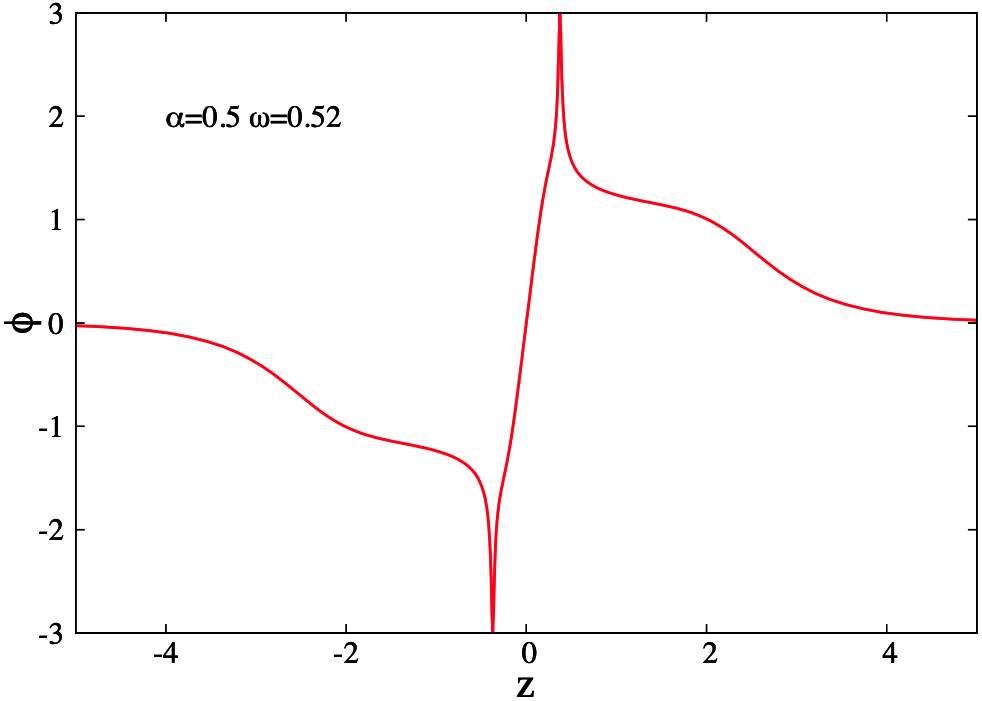}
    \end{center}
    \caption{\small
Profiles on the symmetry axis of the metric function $f$ (left) and the scalar field function $\phi$ of the $k_z=1$
(almost) critical solution on the second branch at $\omega=0.85$ and $\alpha=1$ (upper plots),
and on the third branch at $\omega=0.52$ and $\alpha=0.5$ (lower plots).
Also, $z=r\cos\theta$, with $\theta=0,\pi$.
}
    \lbfig{profiles}
\end{figure}


In fact, the scalar field amplitude
$\phi$
 acquires the shape of two antisymmetric
peak(on)s associated with the locations of the minima of the metric function $f$,
where the name peakons refers to field configurations
with extremely large absolute values of the second derivative at
the maxima \cite{Camassa:1993zz,Lis:2011bc}.
Further numerical investigation of such singular solutions is, unfortunately,
plagued by severe numerical problems.
We remark, that in the case of odd chains the choice of spherical coordinates
and the presence of the major peak(on) at the origin alleviate the numerical
problems considerably.

For smaller $\alpha$,  the coordinate distance $z_\text{d}$ between the
two components of the pair
decreases monotonically towards the limiting solution.
Fig.~\ref{profiles} shows that the closeness of the extrema of the scalar field function
coincides with a very steep rise of the scalar field function at the origin.
The metric function, however, retains only a small peak at the origin. 
Here the numerical grid
allows for better resolution of this extremal behavior,
and thus a closer approach to the limiting solution.
Still, the approach towards the limiting solution is restricted by
numerical accuracy.

The $k_z=3$ quartet represents a bound state of four BSs, located symmetrically
along the symmetry axis.
It may also be viewed as a bound state of two bound pairs of BSs,
since this configuration is not fully symmetric.
The two inner BSs are slightly larger than the two outer BSs,
though there is not too much distinction between the stars
as long as the configuration resides on the fundamental branch.
As seen in Fig.~\ref{comparison-2}, these quartet configurations
share most of the properties with $k_z=1$ pairs.
For the chosen values of the coupling $\alpha$
we find two branches of solutions, the fundamental branch
connected to the perturbative excitations at $\omega \to \omega_\text{max}$,
and the second branch leading to a
limiting solution.
While approaching this limiting solution the outer extrema of the metric and of the scalar field function
become less pronouced, leaving basically the two inner extrema,
which evolve completely analogously to the extrema of the pair
towards a
limiting solution. 
We expect this scenario to represent a generic pattern seen
for all chains with odd $k_z$
(although so far we have checked it for $k_z=1,3,5$ only).

\subsubsection{Odd chains}

As noticed above,
the odd chains always possess a BS
 centered at the origin, with the remaining BSs are
located symmetrically with respect to the origin, on the symmetry axis.
The presence of the central BS constituent suggests  that the $(\omega,M)$-pattern  of the odd chains
could resemble that  found  for a  single BS.
To scrutinize this conjecture, let us consider the branch structure for the
$k_z=2$ triplet of BSs, exhibited in Fig.~\ref{comparison-4}
for four values of the coupling $\alpha$ (with different colors).
Analogously to the even chains, we show
in the two upper panels the scaled ADM mass $M$ (left)
and the scaled charge $Q$ (right).
The  middle panels show
the minimal value $f_\text{min}$ of the metric function (left)
and the maximal value $\phi_{\rm max}$ of the scalar field function (right),
while the lower panel shows
the separation $z_\text{d}$ between the neighboring components of the $k_z=2$ triplet.
Again, all quantities are shown as a function of the frequency $\omega$.

\begin{figure}[p!]
    \begin{center}
        \includegraphics[height=.35\textwidth, trim = 50 20 100 20, clip = true]{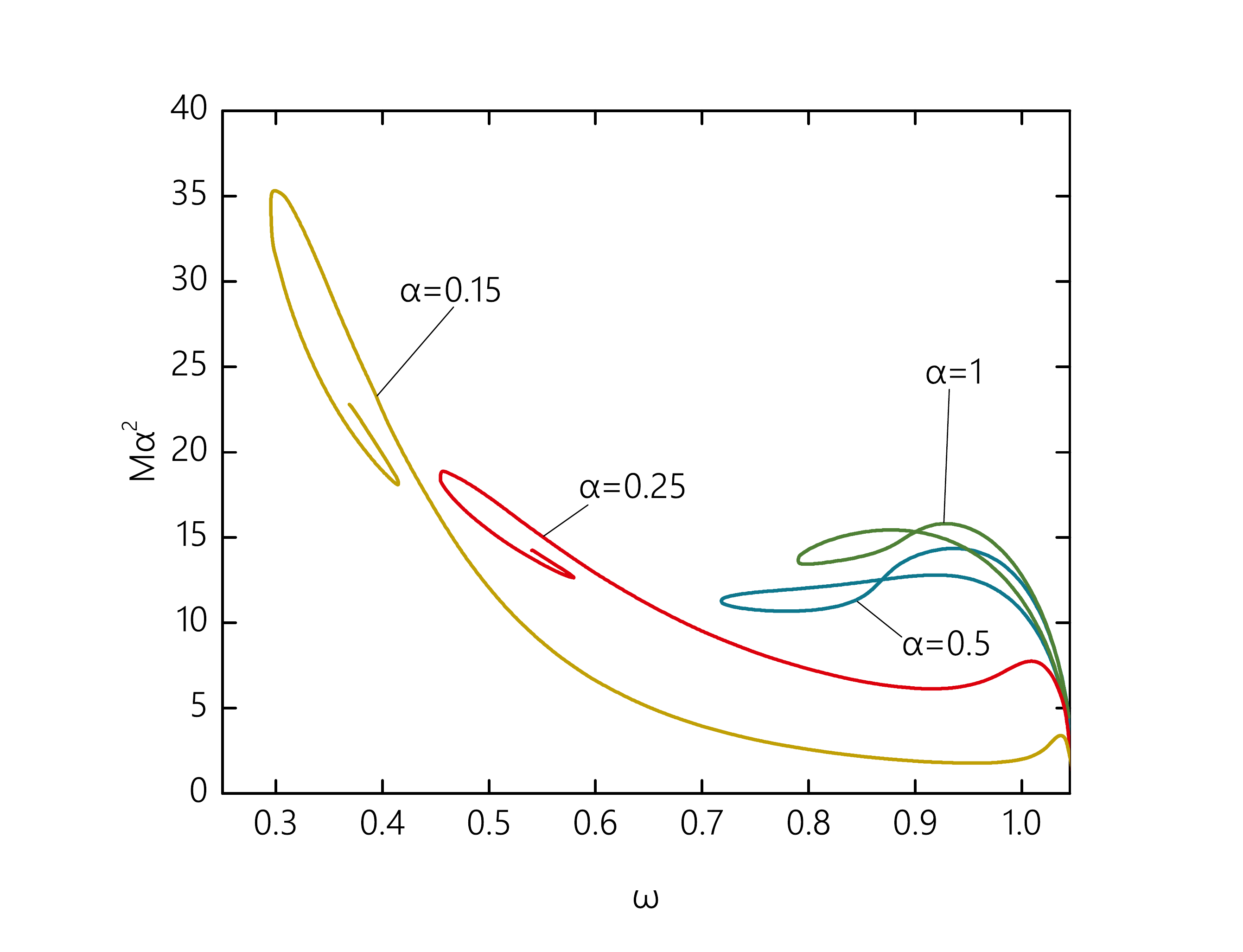}
        \includegraphics[height=.35\textwidth, trim = 50 20 100 20, clip = true]{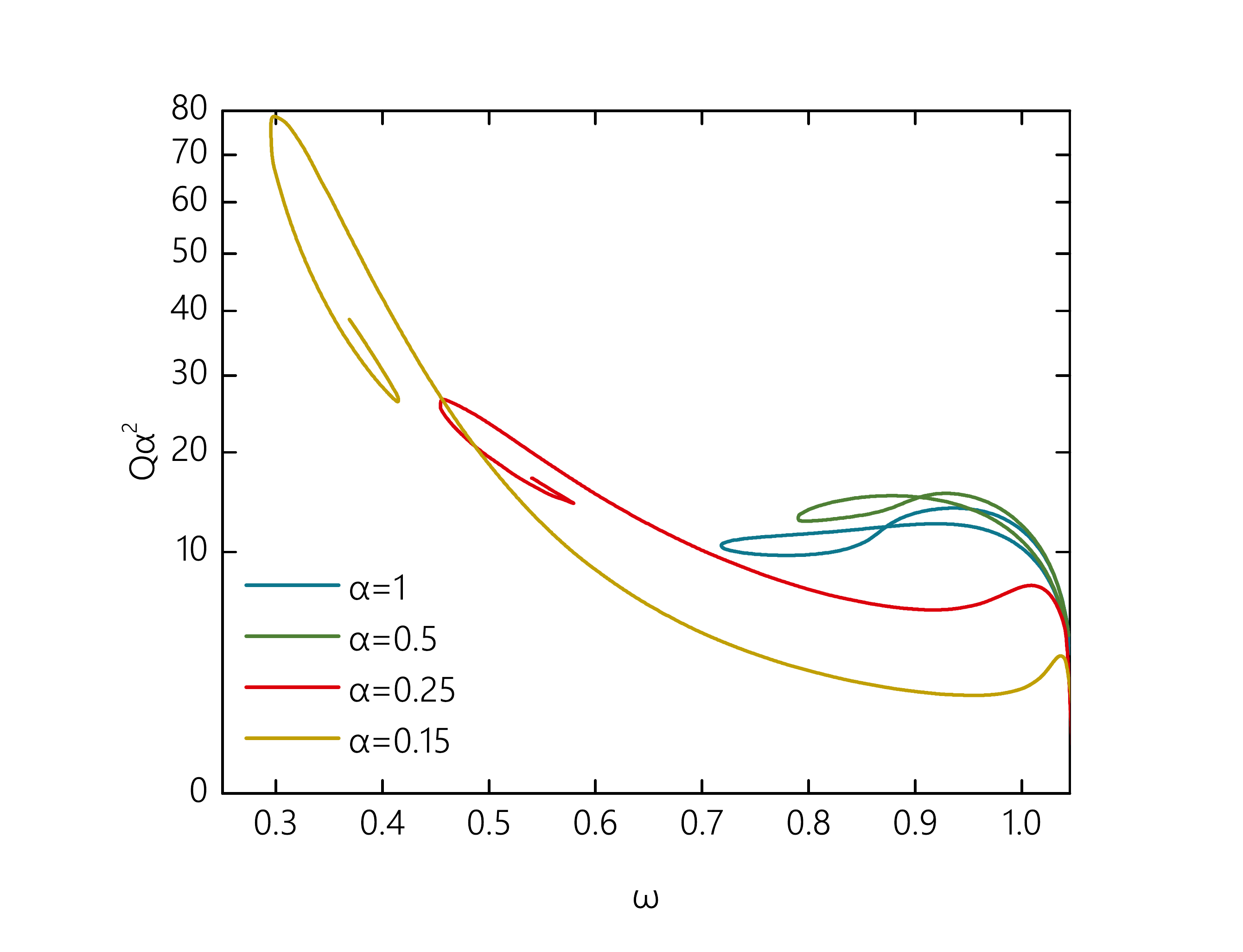}
\includegraphics[height=.35\textwidth, trim = 50 20 100 20, clip = true]{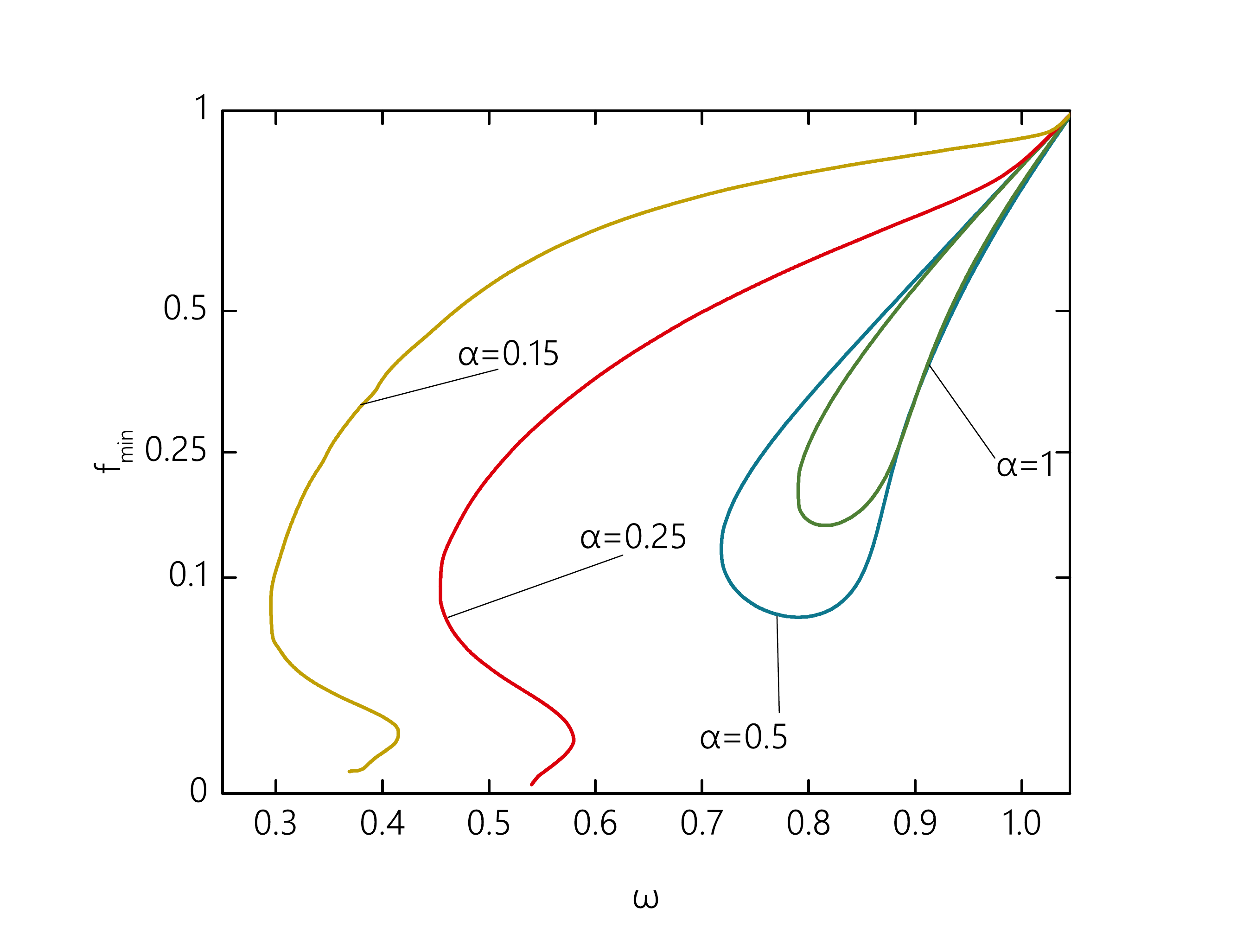}
\includegraphics[height=.35\textwidth, trim = 50 20 100 20, clip = true]{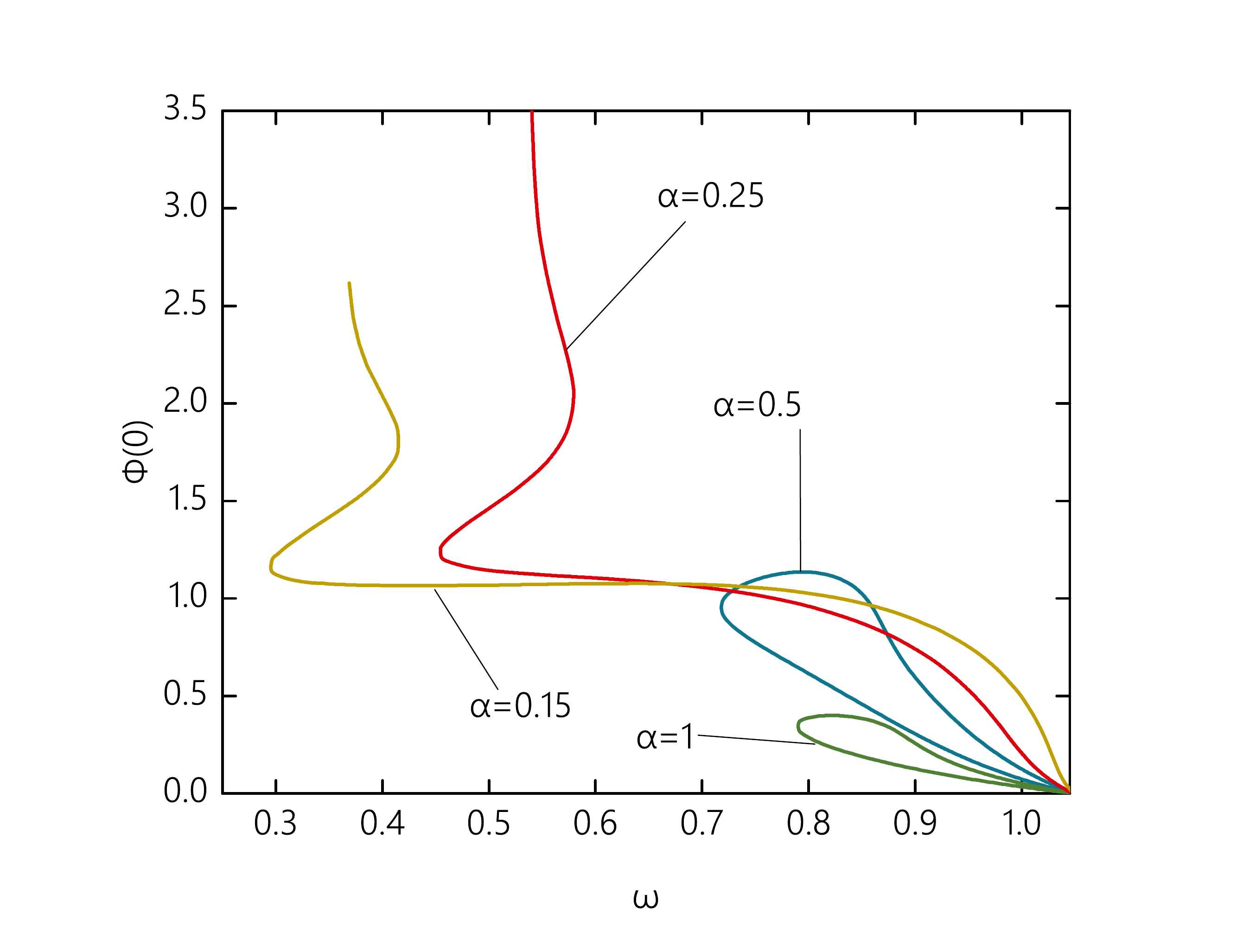}
\includegraphics[height=.35\textwidth, trim = 50 20 100 20, clip = true]{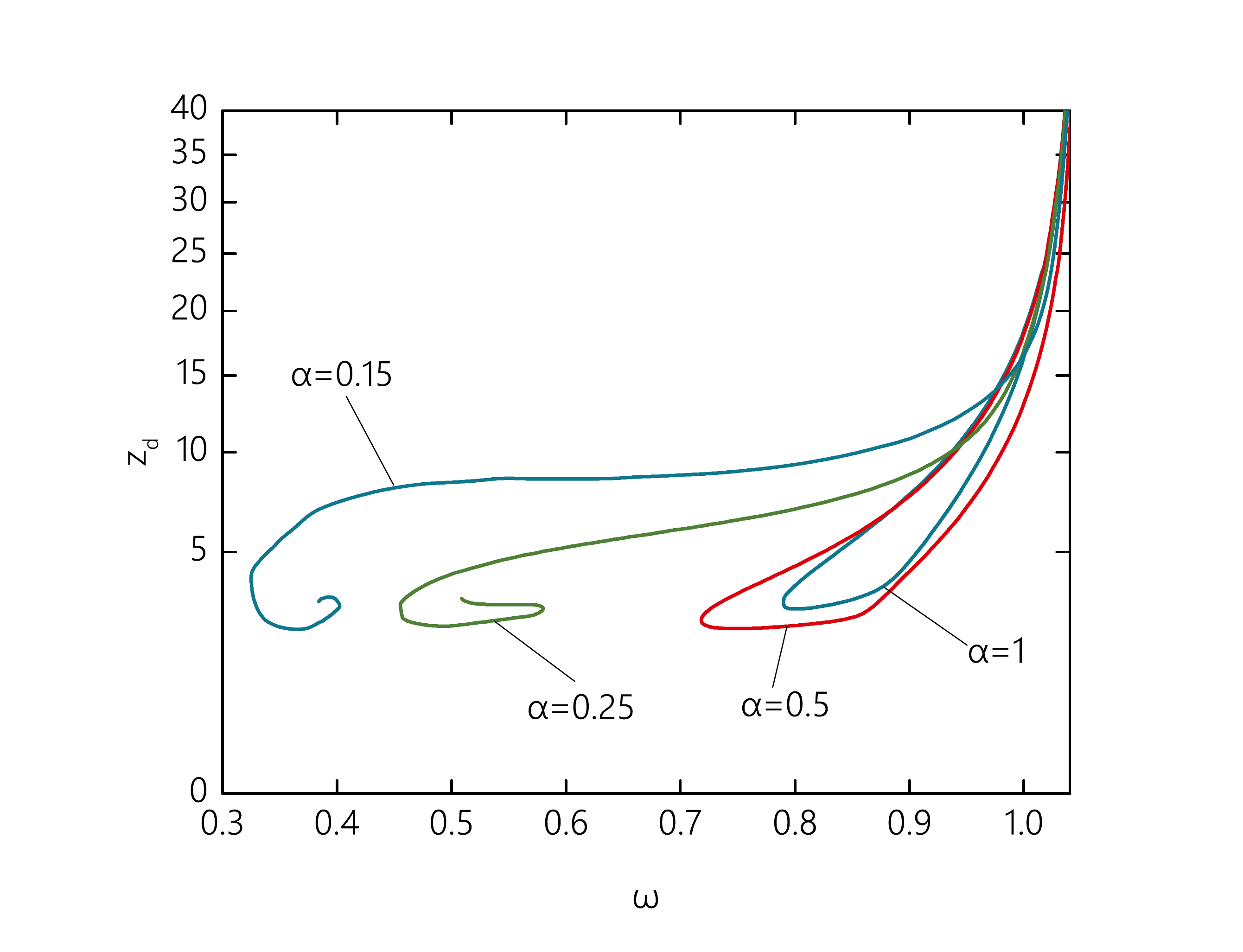}
    \end{center}
    \caption{\small
$k_z=2$ triplet of BSs:
scaled ADM mass $M$ (upper left panel),
scaled charge $Q$ (upper right panel),
minimal value of the metric function $f_\text{min}$ (middle left panel),
maximal value of the scalar field function $\phi_\text{max}$ (middle right panel),
and separation $z_\text{d}$ between neighboring components of the $k_z=2$ triplet (lower panel)
$vs.$ frequency $\omega$ for
several values of the coupling $\alpha$.
Note the quadratic scale for $Q$, $f_\text{min}$ and $z_\text{d}$. }
    \lbfig{comparison-4}
\end{figure}

Let us first consider small values of the coupling $\alpha$.
Then our expectation is borne out, and we observe the triplet forming
spirals for the mass and charge, while damped oscillations for the
extremal values $f_\text{min}$ and $\phi_{\rm max}$. The extremal values
always reside at the center of the configurations, $i.e.$, at the origin.
Also the separation $z_\text{d}$ between the neighboring components of the $k_z=2$ triplet
forms a spiral.
Here, along this spiral, the contribution of the central BS to the full configurations
 becomes dominant, while the outer BSs contributions diminish.


\medskip

When increasing the coupling constant beyond $\alpha \gtrsim 0.45$,
however, the scenario changes completely.
While the configurations still follow a similar  pattern along the fundamental branch,
the further part of the $(\omega,M)$-diagram  does not involve a spiraling/oscillating behavior.
Instead there is a single second branch, that leads all the way back to the
vacuum configuration.
Thus all physical quantities exhibit loops,
beginning and ending at $\omega_\text{max}$.
This may be interpreted as follows.
Along the second branch the configurations again feature a dominant
central BS, but the 'satellite' BSs dissolve into a (sort of) boson shell.
Moreover, the system tends more and more towards spherical symmetry.
Now we recall that
a central BS surrounded by a boson shell is precisely
what constitutes a radially excited spherically symmetric BS with
a single radial node, $n_r=1$.
This suggests that the $k_z=2$ triplet might merge with a $n_r=1$ single BS.

\medskip

Let us therefore compare the $M-\omega$ diagram of the $k_z=2$ triplets at large coupling $\alpha$
with that of the radially excited $n_r=1$ single BSs.
Such a comparison is shown in Fig.~\ref{comparison-4},
where the $k_z=2$ BS triplets are marked by solid curves
and the radially excited $n_r=1$ single BSs by dashed curves
for two values of $\alpha$.
The upper panels show the scaled ADM mass $M$ (left)
and the scaled charge $Q$ (right), while the lower panels show
the minimal value $f_\text{min}$ of the metric function $f$ (left)
and the maximal value $\phi_{\rm max}$ of the scalar field function $\phi$ (right).


\begin{figure}[t!]
\begin{center}
\includegraphics[height=.35\textwidth, trim = 70 20 100 20, clip = true]{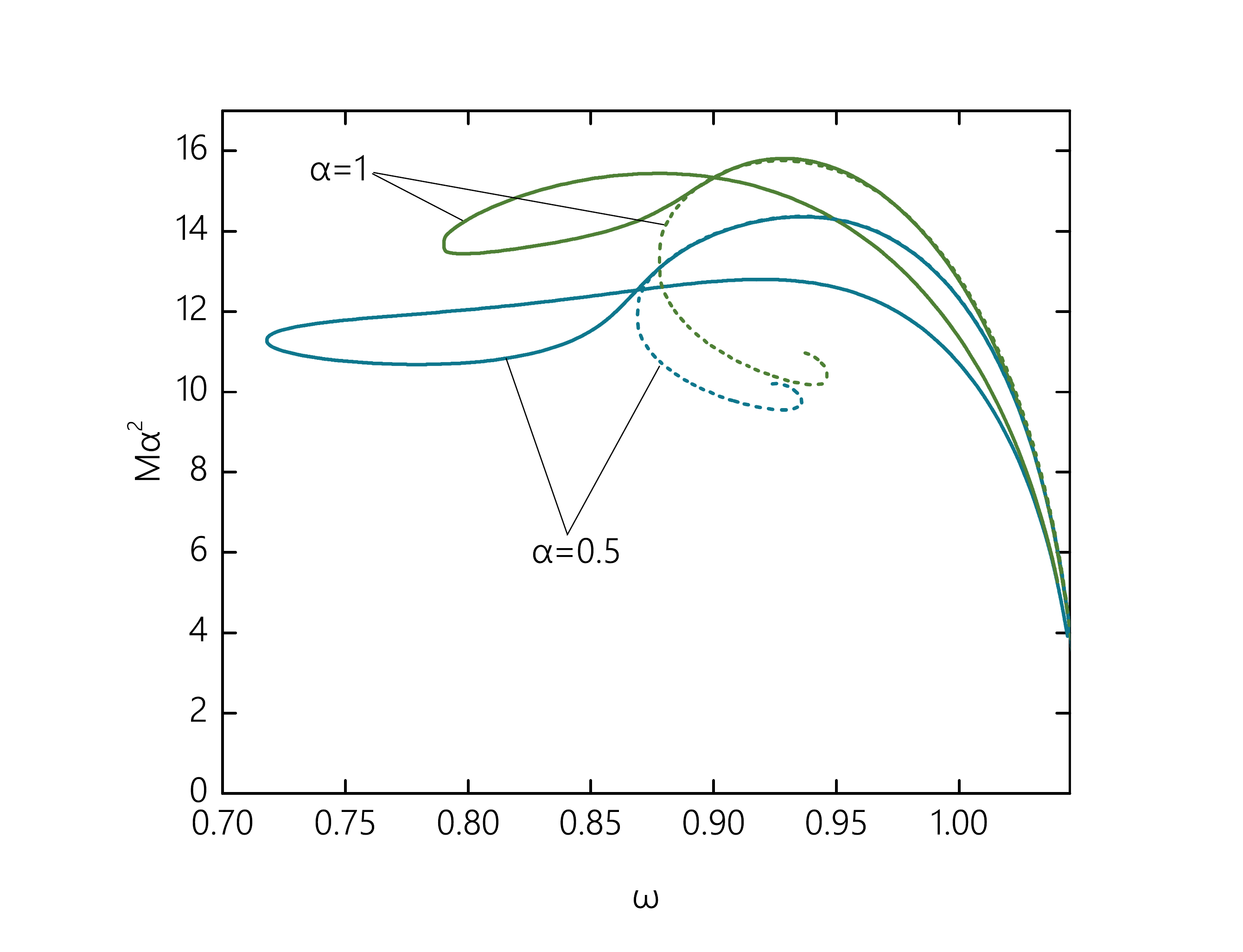}
\includegraphics[height=.35\textwidth, trim = 70 20 100 20, clip = true]{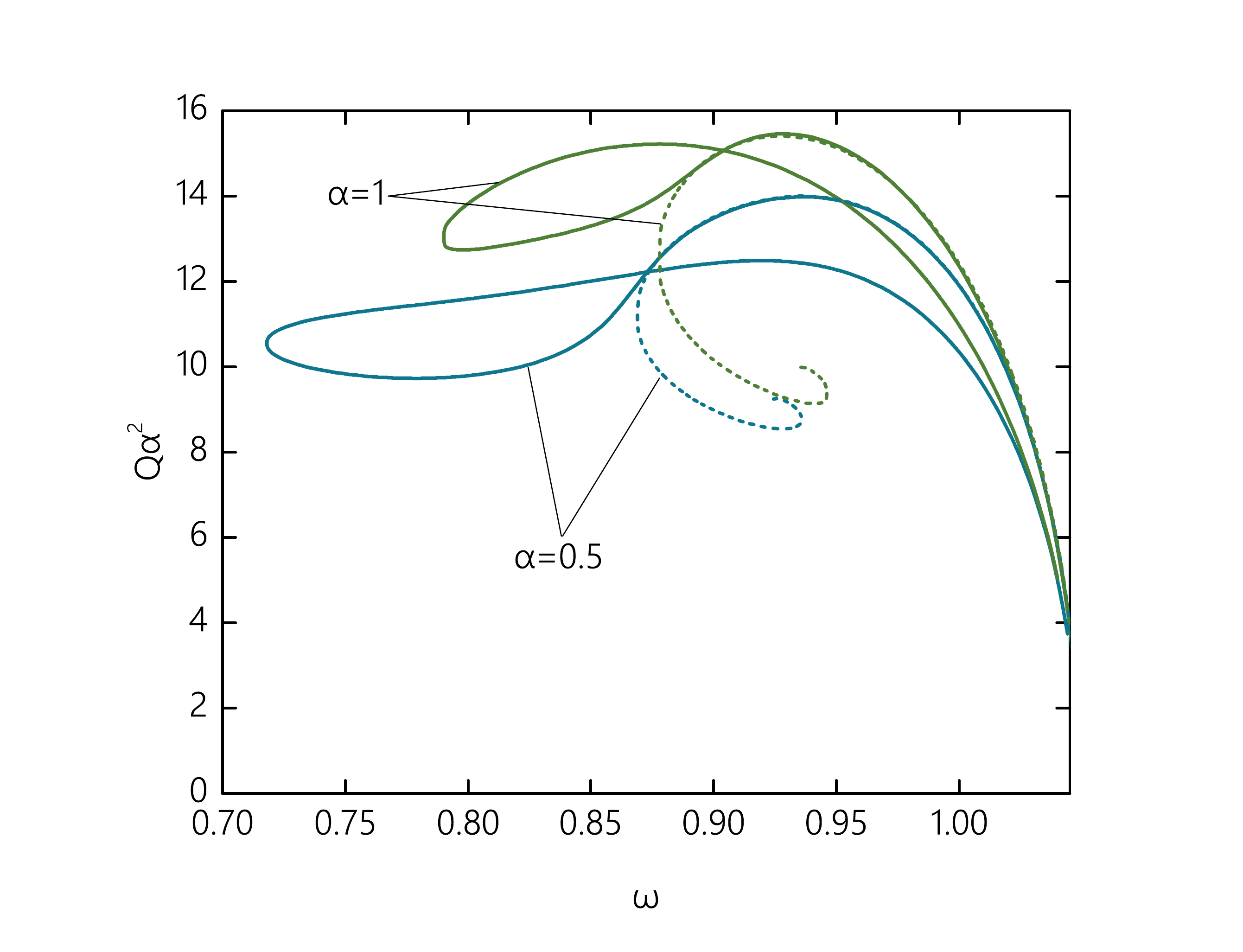}
\includegraphics[height=.35\textwidth, trim = 70 20 100 20, clip = true]{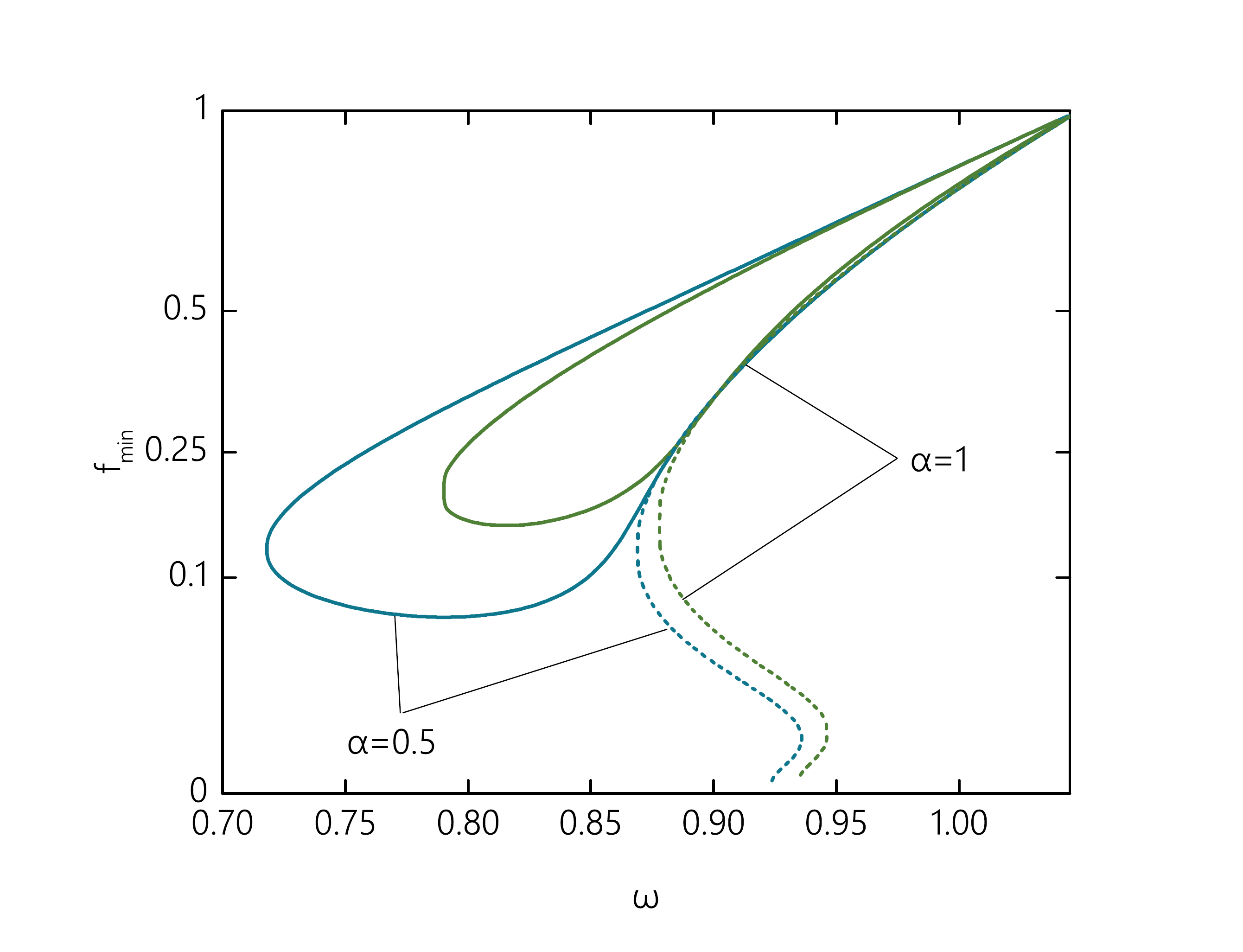}
\includegraphics[height=.35\textwidth, trim = 70 20 100 20, clip = true]{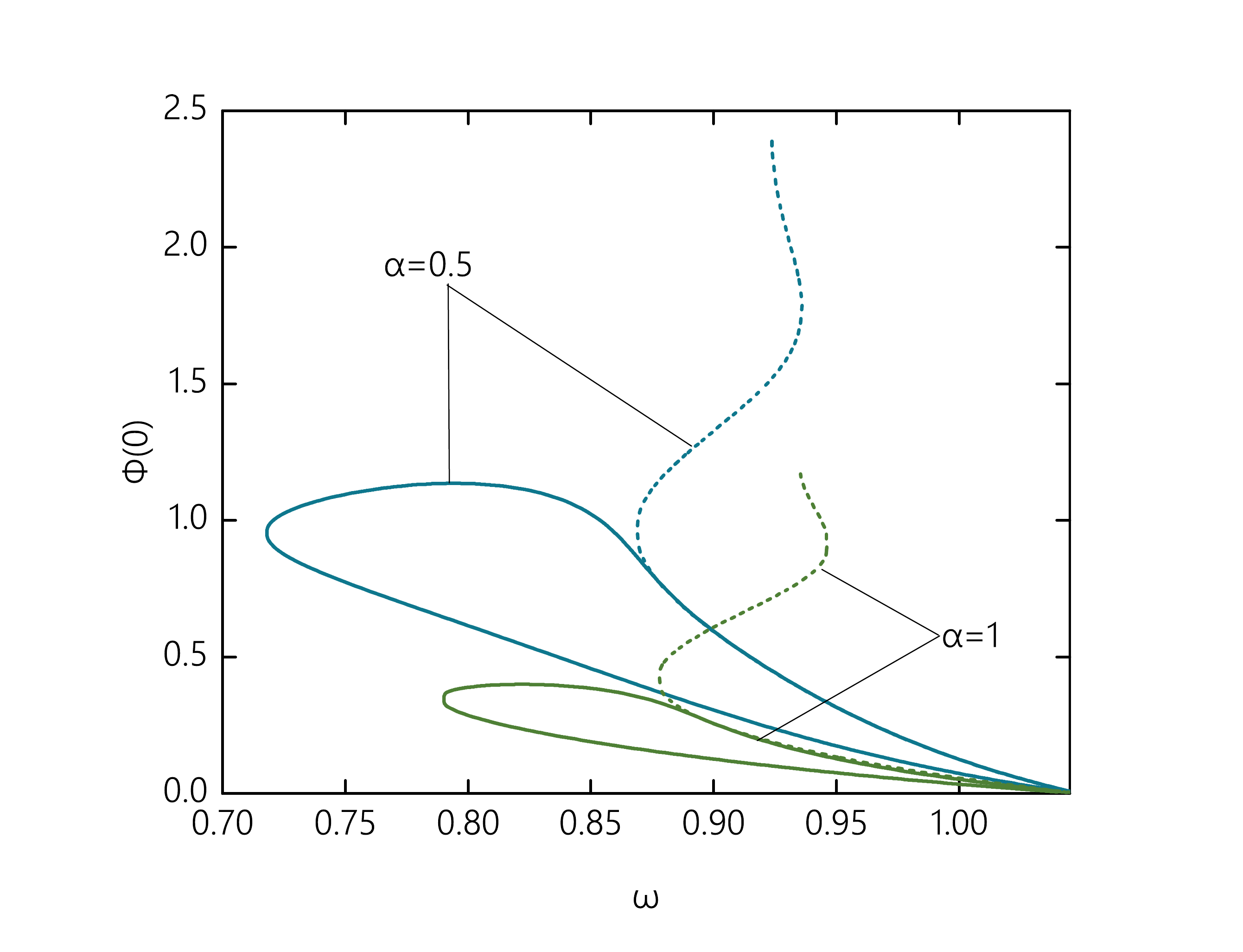}
\end{center}
\caption{\small
    Comparison of the radially excited $n_r=1$ single BSs (dashed curves),
    and $k_z=2$ triplet of BSs (solid curves):
scaled ADM mass $M$ (upper left panel),
scaled charge $Q$ (upper right panel),
minimal value of the metric function $f_\text{min}$ (lower left panel),
maximal value of the scalar field function $\phi_\text{max}$ (lower right panel)
$vs.$ frequency $\omega$ for
two values of the coupling $\alpha$.
Note the quadratic scale for $f_\text{min}$. }
\lbfig{comparison-3}
\end{figure}

These figures are indeed very telling. The radially excited $n_r=1$ single BSs
exhibit the typical curve pattern of spherically symmetric BSs.
They emerge from the vacuum, form the fundamental branch and
end in a spiraling/oscillating pattern.
The $k_z=2$ triplet likewise emerges from the vacuum, forms
a fundamental branch, and a second branch, but this
second branch of the triplet nicely overlaps with the fundamental
branch of the $n_r=1$ single BSs at some critical value
of the frequency $\omega_\text{cr}$.
The overlap happens when mass and charge of the
$n_r=1$ single BSs have already passed their maximal values
and the radially excited stars are descending into the spiral.
 
It is now clear how the domain of existence of odd chains with more constituents should  be, for
sufficiently large values of the coupling $\alpha$.
Let us consider a chain with $k_z$ nodes and thus $k_z+1$ constituents.
Then this chain will feature on its fundamental branch $k_z+1$
more or less equal BSs, located on the symmetry axis.
Subsequently the central BS will start to dominate
while the satellite BSs will dissolve into boson shells.
As the system tends towards spherical symmetry, it will overlap
with a $n_r=k_z/2$ single BS at some critical value
of the frequency.
We have checked this behavior for the $k_z=4$ BS quintet,
which indeed overlaps with the radially excited $n_r=2$ single BS.

\section{Conclusions}

The main purpose of this paper was to report the existence
of a new type of solitonic
configurations for a model with a gravitating
self-interacting complex scalar field.
These configurations represent chains of BSs,
with  $k_z+1$ constituents
located symmetrically along the symmetry axis.
The number $k_z\geqslant 0$ represents the number of nodes on
the symmetry axis of the scalar field amplitude $\phi$.

The chains emerge from the vacuum $\phi=0$ at a maximal value of the boson field frequency $\omega$,
which is given by the field's mass.
For any $k_z$, a fundamental branch
of solutions is found emerging from this vacuum, in a ($\omega,M$)-diagram (with $M$ the ADM mass),
ending at a minimal value of the frequency $\omega_{\rm min}$, whose value is determined by
the self-interaction potential and the gravitational coupling strength $\alpha$.
The subsequent solution curve
depends on the number of
constituents and the coupling $\alpha$.
A single spherical BS has been argued to form an infinite number of branches, leading to spirals or damped oscillations (depending on the quantities considered)
as its limiting configuration is approached.
For even chains we do not see such an endless spiraling/oscillating pattern.
Instead we observe only two to three branches, depending on the
coupling $\alpha$.

As the limiting configuration is approached the even chains retain basically
two of their $k_z+1$ constituents, whose metric function $g_{00}$ exhibits two
sharp peaks, reaching a very small value, while the scalar field features two sharp
opposite extrema located right at the location of these peaks.
The resulting configurations
then feature huge second derivatives of the functions,
which impede further numerical analysis towards the limiting solution.

The odd chains show a similar pattern as the single BS, when
the coupling $\alpha$ is small. This may be interpreted as the central BS dominating the configurations
on the higher branches.
For larger $\alpha$, however, the pattern changes totally,
and the chains overlap on their second branch
with a radially excited spherical single BS with $n_r=k_z/2$ (radial) nodes.
In this case the central dominant BS
will be surrounded by $n_r$ `boson shells'.\footnote{
Boson shells without a central BS
are also known \cite{Kleihaus:2009kr,Kleihaus:2010ep}.}

\medskip

 These solutions can be generalized in various directions.
The most obvious generalization is to include rotation.
For the scalar field this means to include an explicit
harmonic dependence on the azimuthal angle $\varphi$.
The rotating single-BSs were obtained long ago
\cite{Schunck:1996he,Yoshida:1997qf,Ryan:1996nk,Kleihaus:2005me}.
The rotating BSs with odd parity and $k_z=1$,
representing the rotating generalizations of the pair
of BSs, were also discussed in the literature \cite{Kleihaus:2007vk}.
We predict the existence of
rotating generalizations for the triplet and the higher chains discussed in this work.

Single non-rotating BSs cannot be endowed with a black hole at the center;
the no-hair theorems forbid their existence
\cite{Herdeiro:2015waa}.
This result is, however, circumvented
in the presence of spin,
hairy generalizations of the Kerr black hole (BH)
with a complex scalar field
being reported in literature
\cite{Hod:2012px,Herdeiro:2014goa,Herdeiro:2014jaa,Herdeiro:2015gia,Hod:2014baa,Benone:2014ssa,Herdeiro:2014pka,Herdeiro:2014ima}.
These hairy BHs  obey a synchronization
condition relating the angular velocity of the event horizon
and the boson field frequency.
Most studies so far considered only an even parity scalar field,
see $e.g.$
\cite{Kleihaus:2015iea,Herdeiro:2015tia,Herdeiro:2015kha,Herdeiro:2016tmi,Brihaye:2016vkv,Hod:2017kpt,Herdeiro:2017oyt,Herdeiro:2018daq,Herdeiro:2018djx,Wang:2018xhw,Delgado:2019prc,Kunz:2019bhm,Kunz:2019sgn}.
Synchronized hairy BHs with an  odd parity scalar field were obtained
in \cite{Wang:2018xhw,Kunz:2019bhm,Kunz:2019sgn}.
While these solutions represent only the simplest type of generalizations, containing a single black hole
at the center of configurations with one (parity even) constituent or two (parity odd) constituents
one can easily image to put a black hole either at the center of rotating configurations with more than two
components, or to put a black hole at the center of each of the components along the symmetry axis.
Such configurations should correspond to hairy double Kerr solutions or hairy multi-Kerr solutions.
It would be interesting to see, whether the presence of the scalar hair can regularize such solutions,
so that no conical singularity would be needed to hold them in equilibrium.

Along similar lines, but replacing rotation by an electric charge,
one could investigate chains of electromagnetically
charged BSs, generalizing the results for a single charged BSs in Einstein-Klein-Gordon-Maxwell model
\cite{Jetzer:1989av,Kleihaus:2009kr,Herdeiro:2020xmb}.
Some results in this direction were reported in the recent work
\cite{Loiko:2020htk},
where (flat space) $Q$-chains were constructed in a model with a $U(1)$ gauged scalar field,
for a particular choice of the constants in the potential  (\ref{pot}).
Gravitating generalizations of these solutions should also exist, sharing some of the properties
of the (ungauged) BS chains in this work.
In this context,
 we mention the recent results in
\cite{Herdeiro:2020xmb,Hong:2020miv},
showing that
the no-hair theorem in
\cite{Mayo:1996mv}
 does not apply to a single charged static
BS
in a model with a $Q$-ball type potential  (\ref{pot}), with the existence of BH generalizations.
For chains of charged BSs, it would be again of particular interest to see  whether
they would support regular static multi-BH solutions.

\section*{Acknowledgements}

We gratefully acknowledge support by the DFG funded
Research Training Group 1620 ``Models of Gravity''
and by the DAAD.
We would also like to acknowledge networking support by the
COST Actions CA16104 and CA15117.
Ya.S. gratefully acknowledges the support by
the Ministry of Sceince and High Education of
Russian Federation, project FEWF-2020-0003. and by the BLTP JINR Heisenberg-Landau program 2020.
This  work  is also supported  by  the Center  for  Research  and  Development  in  Mathematics  and  Applications  (CIDMA)  through  the Portuguese Foundation for Science and Technology (FCT - Fundacao para a Ci\^encia e a Tecnologia), references UIDB/04106/2020 and UIDP/04106/2020 and by national funds (OE), through FCT, I.P., in the scope of the framework contract foreseen in the numbers 4, 5 and 6 of the article 23, of the Decree-Law 57/2016, of August 29, changed by Law 57/2017, of July 19.  We acknowledge support  from  the  projects  PTDC/FIS-OUT/28407/2017,  CERN/FIS-PAR/0027/2019 and PTDC/FIS-AST/3041/2020.   This work has further been supported by the European Union’s Horizon 2020 research and innovation (RISE) programme H2020-MSCA-RISE-2017 Grant No. FunFiCO-777740.
The authors would like to acknowledge networking support by the COST Action CA16104.


\end{document}